\documentclass[12pt]{article}

\usepackage{amsmath}
\usepackage{amssymb}
\usepackage{color}
\usepackage{hyperref}
\usepackage{subfigure}
\usepackage{graphicx}
\newcommand{\B}[1]{\mbox{\boldmath${#1}$\unboldmath}}
\newcommand{\R}{\mathbf{R}}

\definecolor{DarkGreen}{RGB}{1, 50, 32}

\begin{document}


\title{Nonlinear propagating localized modes in a 2D hexagonal
    crystal lattice}

  \author{J. Bajars$^{(1,2)}$, J. C. Eilbeck$^{(1,3)}$, and
    B. Leimkuhler$^{(1,2)}$\\
    $^{(1)}$Maxwell Institute, \\
    $^{(2)}$School of Mathematics, University of Edinburgh\\ James
    Clerk Maxwell Building, The King's Buildings,\\ Mayfield Road,
    Edinburgh EH9 3JZ, UK,
    \\$^{(3)}$Department of Mathematics, Heriot-Watt University,\\
    Riccarton, Edinburgh EH14 4AS, UK}



\maketitle

\begin{abstract}
  In this paper we consider a 2D hexagonal crystal lattice model first
  proposed by Mar\'in, Eilbeck and Russell in 1998.  We perform a
  detailed numerical study of nonlinear propagating localized modes,
  that is, propagating discrete breathers and kinks.  The original
  model is extended to allow for arbitrary atomic interactions, and to
  allow atoms to travel out of the unit cell.  A new on-site potential
  is considered with a periodic smooth function with hexagonal
  symmetry.  We are able to confirm the existence of long-lived
  propagating discrete breathers.  Our simulations show that, as they
  evolve, breathers appear to localize in frequency space, i.e.\ the
  energy moves from sidebands to a main frequency band.  Our numerical
  findings contribute to the open question of whether exact moving
  breather solutions exist in 2D hexagonal layers in physical crystal
  lattices.
\end{abstract}




\section{Introduction}\label{sec:Intro}
This article examines propagating nonlinear localized modes in
crystalline materials. These modes are commonly referred to as
propagating discrete breathers, i.e.~waves contained within a
bell-shaped envelope exhibiting internal oscillations of frequencies
outside the phonon band. There is strong interest in such breather
solutions, as they provide possible mechanisms underpinning physical
phenomena, for example the formation of long decorated dark lines in
muscovite mica \cite{ProcMica14,Ru88,Ru91,RuCo95,RuEi11}, a possible
mechanism for high temperature superconductivity
\cite{RuCo96,MaEiRu01}, and the development of next generation plasma
fusion reactors \cite{RuEi07}. Laboratory and numerical experiments
provide evidence for such coherent localized phenomena
\cite{SeAkRu00,RuEi07,SwEtAl99,DoCuEiRu11,MaEiRu98,MaEiRu01,MaEiRu00,IkDoFeKa07}.

The existence of dark lines in muscovite mica crystals was first
highlighted by Russell some twenty years ago \cite{RuCo95}.  Although
some of these lines were thought to be formed by cosmic rays, the fact
that many of the lines follow crystallographic axes was puzzling.
Russell was unable to find a suitable linear theory for such phenomena
and suggested the possibility of nonlinear localized modes.  He called
these ``quodons'' as they appeared to be connected to a symmetry
feature of the axes which he called quasi-one-dimensionality, for
which displacement of an atom along the axis direction was met by a
force acting along the same line (technically this is $C_2$ symmetry).
In this paper we use the term quasi-one-dimensional to refer both to
this type of symmetry and to the fact that the observed mobile pulses
seem to be highly localized along one of the crystallographic axes.
The two effects are believed to be related \cite{MaEiRu98}, although the
exact mechanism is not clear.  

The active component in the mica case seems to be the 2D hexagonal
layer of K atoms sandwiched between two relatively rigid silicate
layers.  Such a symmetry feature may be also associated with many of
the materials having high $T_{c}$ superconductivity properties
\cite{RuCo96}, although in this case the underlying 2D layers have a
higher cubic rather than hexagonal symmetry.  More tenuous suggestions
that localized modes could lead to enhanced fusion rates in deuterated
crystals have also been put forward \cite{RuEi07}.

More generally, there is increasing interest in single 2D hexagonal
crystals such as graphene \cite{ge09} and other layered structures
that could be built from the two-dimensional atomic crystals based on
graphene geometries \cite{GeGr13}.  An open question is what role both
stationary and mobile localized modes can play in such structures.

Due to advances in computer power and better understanding of
molecular dynamics algorithms \cite{AlTi89,LeRe05}, we are now well
equipped for numerical study of discrete breathers in higher
dimensional dynamical lattices.  There is a good theoretical and
numerical understanding of existence of stationary discrete breathers
\cite{FlGo08,Au06,MaAu94}, that is, spatially localized time-periodic
excitations.  The same cannot be said about mobile discrete breathers
in 1D and higher dimensional dynamical lattices
\cite{FlGo08,Au06,FlKl99,AuCr98,MaSe02}.  There are still open
theoretical questions regarding the existence of propagating discrete
breathers in general nonintegrable lattices.  There are a few
exceptions such as the Ablowitz-Ladik chain \cite{AbLa76}, which is an
integrable model.  Thus we must rely on numerical studies of
propagating localized modes, the main focus of this paper.  In
addition, we propose here a 2D model with a lower level of complexity
for future analytical investigations.

In their work, Mar\'in et al.\ \cite{MaEiRu98,MaEiRu00} showed
numerically for the first time the existence of propagating localized
modes (discrete breathers) in a 2D dynamical hexagonal nonlinear
lattice.  They extended their results to a 2D cubic lattice in
\cite{MaEiRu01}.  Their lattices were subject to a nearest neighbour
anharmonic interparticle interaction potential and an on-site
potential.  Their model represents a 2D nonlinear lattice when
embedded in a surrounding 3D lattice and as such can be thought of as
a 2D layer model of a 3D layered crystal lattice.  Examples of such
crystals are cuprates, the copper-oxide based high temperature
superconductors, with cubic symmetry, and muscovite mica, a potassium
based silicate insulators, with hexagonal symmetry.  The study
\cite{MaEiRu98} was limited by available computer power and hence only
to models of $16^2$ and $32^2$ lattices sizes (i.e. $<1000$ lattice
sites) with periodic boundary conditions.  This numerical study
confirmed the existence of propagating highly localized
quasi-one-dimensional discrete breathers propagating in
crystallographic directions.  The quasi-one-dimensional nature of the
discrete breathers suggests that they may exist in crystals containing
any 1D chains with $C_{2}$ symmetry.

The present paper explores the hexagonal model in much greater detail,
using a much larger computational domain to simulate a system of up to
$3\times10^5$ lattice sites.  Using somewhat smaller 2D domains but
with periodic boundary conditions, we are able to track a breather
traversing one million lattice sites.  Our study shows frequency
sharpening effects in both 1D and 2D, which were not observed in the
paper of Marin et al.\ \cite{MaEiRu98}. Moreover, we obtain a better
understanding of the 2D qualitative nature of these
quasi-one-dimensional propagating breathers.

For the theory stated above to be a proper representation of physical
reality, discrete breathers must travel long distances, i.e.~one or
more millimeters, if we are to associate them with the creation of dark lines
in mica and to give numerical support for sputtering experiment
carried out by Russell et al.\ \cite{RuEi07}.  In this experiment, a
specimen of muscovite mica of size $\sim1$mm.\ in thickness and
$\sim$7mm.\ across the $(001)$-face was subject to low energy alpha
particle bombardment at one end of the crystal. The experiment showed
that particles were emitted at the opposite face of the specimen in
the crystallographic directions of the potassium layer of muscovite.
The potassium layer is thought to be the layer where discrete
breathers could have propagated \cite{RuEi11}.  In general discrete
breathers are not {\em a priori} expected to be long-lived since the
lattice models considered are likely to be nonintegrable.  In
addition, their lifespan is subject to interactions with defects and
with the phonon background, which may be viewed as thermal noise.  The
experiment by Russell et al.\ \cite{RuEi07} showed the transport of
energy over more than $10^7$ lattice sites at about $300^\circ$K (room
temperature).  An obvious challenge is to demonstrate the existence of
long-lived propagating discrete breather solutions using a theoretical
model to understand their role in the transport of energy in
sputtering experiments and formation of the dark lines in mica.

The simulations of Mar\'in et al.\ \cite{MaEiRu98} were done on small
lattices with periodic boundary conditions: the longest distance of
breather travel was reported to be $\le{10}^4$ lattice sites
(i.e.\ traversing the periodic lattice many times) before
the wave collapsed.  Their results suggest that the discrete breathers
are sensitive to small scale interactions with phonon background,
i.e.~thermal noise may turn propagating modes into stationary ones, or
scatter their energy into the lattice, and thus lifetimes were
insufficient to support the experimental results of Russell et al
\cite{RuEi07}.  On the other hand this does not preclude the
existence of long-lived propagating breather solutions when the right
conditions are met, since the study by Cretegny et al.\
\cite{CrAuFl98} showed that phonons may turn stationary breathers into
a mobile one.  The paper \cite{MaEiRu98} used a lattice configuration
in its dynamical equilibrium state for the initial conditions, while
additionally exciting three atoms in one of the crystallographic
direction with positive-negative-positive or vice versa initial
velocity conditions.  With these initial conditions they were able to
create a propagating breather solution together with low amplitude
phonons which spread into the domain and continued to interact with
the propagating breather.  Thus these interactions could be
responsible for the collapse of propagating discrete breather.  

The model in \cite{MaEiRu98} was highly restricted in that it
incorporated only nearest neighbour interactions between potassium
atoms, with the atoms confined to their unit cells.  Thus they were
not able to study kink solutions in a 2D hexagonal lattice. The
objective of this paper is to eliminate this constraint, to allow
short and long range interactions between potassium atoms and perform
a conceptual numerical study of long-lived breather solutions.  The
current approach also allows us to study kink solutions in a 2D
hexagonal lattice.

The paper is organised in the following way.  In Section
\ref{sec:MathMod} we consider in detail the theoretical model we use
in our study.  We derive a dimensionless set of equations in Sec.\
\ref{sec:DimSyst}, and in Sec.\ \ref{sec:LinEq} we investigate the
linearised system and derive the linearised dispersion relation.  In
Sec.\ \ref{sec:NumSim} we report on a number of numerical simulations
of long-lived breathers in our model system.  Section \ref{sec:Kinks}
is devoted to a brief discussion of simulations of kink solutions,
which do not appear to travel long distances in the present model.
Some mathematical details are presented in two appendices.

\section{Mathematical model}\label{sec:MathMod}
In this section we describe a 2D mathematical K-K sheet
layer model of muscovite mica crystal.  We model the potassium layer of
mica of $N$ potassium atoms by classical Hamiltonian dynamics with the
Hamiltonian:
\begin{align}\label{Hamilt}
\begin{split}
  H &= K + V + U \\
  &= \sum_{n=1}^{N} \biggl( \frac{1}{2} m \|\dot{\B{r}}_{n}\|^2 +
    U(\B{r}_{n}) + \frac{1}{2} \sum_{\substack{n'=1,\\n'\neq{n}}}^{N}
    V(\|\B{r}_{n}-\B{r}_{n'}\|) \biggr),
\end{split}
\end{align}
where $K$ is the kinetic energy, $U$ is
the on-site potential energy encompassing forces from the silicate
layers of atoms above and below the potassium K-K sheet,
and $V$ is the radial interaction potential between the potassium
atoms. In the Hamiltonian (\ref{Hamilt}), $\B{r}_{n}\in\R^2$ is the 2D
position vector of the $n^{th}$ potassium atom with mass $m$, and
$\dot{\B{r}}_{n}$ is its time derivative.  The symbol $\|\B{u}\|$ refers
to the Euclidean two-norm of a vector $\B{u}$, i.e.\ its length.

\subsection{Forces between crystal layers}
\label{sec:OnSite}
\begin{figure}[ht]
\centering 
\subfigure[]{\label{fig:OnSiteA} \includegraphics[scale=0.42]
{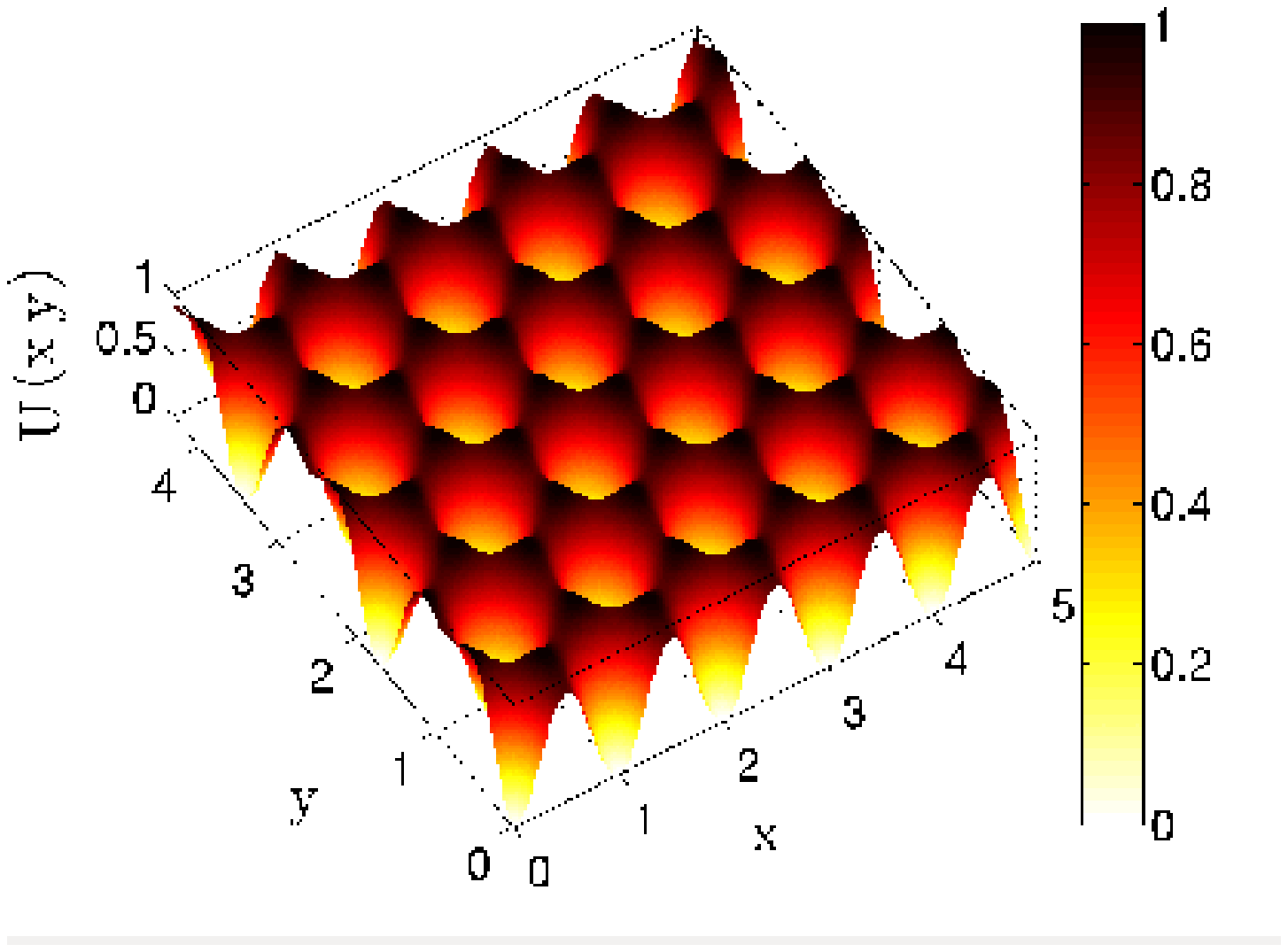}}
\subfigure[]{\label{fig:OnSiteB} \includegraphics[scale=0.33]
{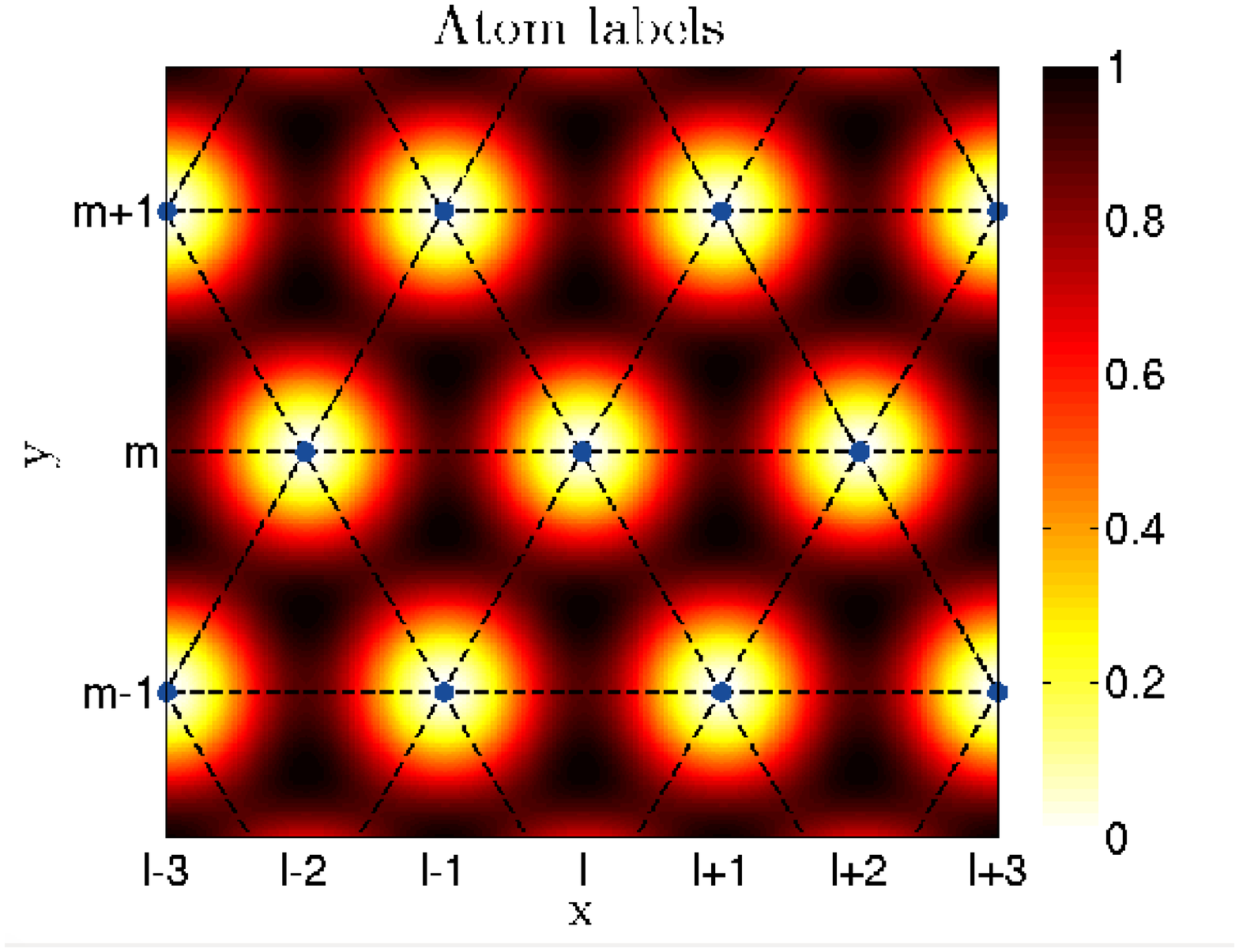}} 
\caption{Smooth periodic on-site potential function with hexagonal
  symmetry in $(x,y)$ coordinates. (a) on-site potential with sixteen
  sites, $\sigma=1$ and $U_{0}=1$. (b) configuration of potassium
  atoms in dynamical equilibrium states with discrete labels $(l,m)$
  in $(x,y)$ coordinates.}\label{fig:OnSite}
\end{figure}

The potassium K-K sheet of muscovite mica crystal is compactly
sandwiched between rigid layers of silicon-oxygen tetrahedra which
enforces hexagonal lattice symmetry on potassium atoms \cite{RuCo95}.
Mar\'in et al.\ \cite{MaEiRu98} considered the rigid silicon-oxygen
layer approximation and anharmonic interaction Morse forces between
free potassium and fixed oxygen atoms to obtain an on-site force as a
superposition of these forces for each site.  In this paper we adopt a
simpler approach and consider a periodic smooth on-site potential with
hexagonal symmetry from \cite{YaDuYaChLi11}.  This can be thought of
as a generalization of a discrete 1D sine-Gordon lattice to a two
dimensions with hexagonal symmetry.  The on-site potential function
resembles an egg-box carton and can be written as
\begin{align}\label{OnSiteFunc}
\begin{split}
  U(x,y) = &\frac{2}{3} U_{0} \biggl(1-\frac{1}{3} \biggl( \cos{\biggl(
          \frac{4\pi{y}}{\sqrt{3}\sigma} \biggr)} \biggr. \biggr. \\
  & \qquad + \biggl. \biggl. \cos{ \biggl( \frac{2\pi(\sqrt{3}x-y)}
{\sqrt{3}\sigma}\biggr)} + \cos{
        \biggl( \frac{2\pi(\sqrt{3}x+y)}{\sqrt{3}\sigma} \biggr) }
    \biggr) \biggr),
\end{split}
\end{align}
where $\sigma$ is the lattice constant, i.e.~the equilibrium distance
between potassium atoms, and $U_{0}>0$ is the maximal value of the
on-site potential. Note that a simple product of cosine functions
would not provide the required hexagonal symmetry.  In Figure
\ref{fig:OnSiteA}, we plot the on-site potential function
(\ref{OnSiteFunc}) with sixteen sites, $\sigma=1$ and $U_{0}=1$. In
Figure \ref{fig:OnSiteB} we show potassium atoms in their dynamical
equilibrium states together with their labels in $(x,y)$ coordinates.
We will adopt these labels and notation in Secs.\ \ref{sec:LinEq} and
\ref{sec:NumSim}.

For further reference and analysis we write down the harmonic
approximation to an on-site potential well
\begin{align}\label{OnSiteFuncAppr}
  U_{h}(x,y) = \frac{16\pi^2 U_{0}}{18\sigma^2} \biggl( (x-x_{0})^2 +
    (y-y_{0})^2 \biggr),
\end{align}
where $(x_{0},y_{0})$ are any local minima, equilibrium states, of the
on-site potential (\ref{OnSiteFunc}), see Fig.\ \ref{fig:OnSite}.

For small atomic displacements, each potassium atom K will remain in
each particular site.  This was imposed as a global constraint in
\cite{MaEiRu98} with nearest neighbour interactions.  In our approach we
can allow atoms to move from one site to another.  In addition, the
on-site potential (\ref{OnSiteFunc}) provides a simpler
implementation, since function (\ref{OnSiteFunc}) is periodic and
defined on all of $\R^2$.  In the 1D approximation, i.e.~$y=$constant,
this on-site potential (\ref{OnSiteFunc}) reduces to the cosine
function which is the on-site potential of the discrete sine-Gordon
equation and a periodic potential of the 1D model considered in
\cite{DoCuEiRu11}.  The model in \cite{DoCuEiRu11} could be thought as
a 1D approximation of the 2D model (\ref{Hamilt}).  To see that,
without loss of generality, consider $y=0$ in the equation
(\ref{OnSiteFunc}) which leads to the cosine function in the 1D
approximation.  The same holds true for other two crystallographic
lattice directions.  The hexagonal lattice, as demonstrated in Fig.\
\ref{fig:OnSiteB}, has three crystallographic lattice directions which
can be prescribed by the direction cosine vectors: $(1,0)^{T}$ and
$(1/2,\pm\sqrt{3}/2)^{T}$.

\subsection{Nonlinear interaction forces}\label{sec:InterForces}
In this section we describe an empirical interaction potential to
model the atomic interactions of potassium atoms in the K-K
sheet of mica.  Essentially, from a modelling point of view, we are
concerned with anharmonic radial interaction potentials
$V(r)=V(\epsilon,\sigma,r)$ parametrized by $\epsilon>0$, the depth of
the potential well, i.e.~$V(\epsilon,\sigma,\sigma)=-\epsilon$, and
$\sigma>0$, the equilibrium distance,
i.e.~$\partial_{r}V(\epsilon,\sigma,\sigma) = 0$. In addition, we
require that $\partial_{rr}V(\epsilon,\sigma,\sigma) > 0$ and
$V(\epsilon,\sigma,r)$ are monotonically increasing functions for
$r<\sigma$ and $r>\sigma$ such that
\begin{equation}\label{Properties}
 \lim_{r\to\infty}V(\epsilon,\sigma,r) = 0, \quad 
\lim_{r\to\infty}\partial_{r}V(\epsilon,\sigma,r)=0.
\end{equation}
As a first step to understanding the properties of the crystalline
solids in this model, it is natural to consider the short-ranged
scaled Lennard-Jones interaction potential
\begin{equation}\label{LJpot}
 V_{LJ}(r) = \epsilon \left( \left( \frac{\sigma}{r} \right)^{12} 
- 2 \left( \frac{\sigma}{r} \right)^{6} \right),
\end{equation}
where $\sigma$ coincide with the lattice constant in the on-site
potential (\ref{OnSiteFunc}), $r:=r_{n,n'}=||\B{r}_{n}-\B{r}_{n'}||$
for all $n,n'=1,\dots,{N}$ and $n\neq{n'}$. Recall that the term
$r^{-6}$ describes long range attractive van der Waals force and the
term $r^{-12}$ models Pauli short range repulsive forces. Other
possible models are the Morse potential and the Buckingham potential,
among others.

The Lennard-Jones potential (\ref{LJpot}) has the asymptotic
properties (\ref{Properties}).  To increase the efficiency of the
numerical computations, and to provide a suitable model for nearest
neighbour interactions, we introduce an additional parameter in the
potential (\ref{LJpot}), that is, a cut-off radius $r_{c}$.  In this
paper we are concerned with a close range interaction model,
i.e.~$r_{c}=\sqrt{3}\sigma$, which resembles but is not restricted to
the fixed nearest neighbour interaction model.

We compared our numerical results to longer ranged interaction
simulations, that is, with $r_{c}=2\sigma$ and $r_{c}=3\sigma$, and
did not observe any qualitative differences in our results.  We
attribute this to the asymptotic properties (\ref{Properties}) of the
Lennard-Jones potential (\ref{LJpot}).

To incorporate the cut-off radius, we set potential and forces to zero
for all atomic distances larger than $r_{c}$.  For smooth cut-off
computations we proceed in a similar manner as presented in
\cite{StFo73}.  Instead of only two polynomial terms, we add five
additional even order polynomial terms to the interaction potential
$V(r)$, i.e.
\begin{align}\label{CutPot}
 V_{cut}(r) = \left\{
  \begin{array}{l l}
    \displaystyle V(r) + \epsilon \sum_{j=0}^{4} A_{j} 
\left( \frac{r}{r_{c}} \right)^{2j} , & \quad 0 < r \leq r_{c}, \\
    0, & \quad \mbox{otherwise},
  \end{array}
\right.
\end{align}
where the cut-off dimensionless coefficients $A_{j}\to{0}$ when
$r_{c}\to{\infty}$ and are determined from the following five
conditions:
\begin{align}\label{Cond}
\begin{split}
  V_{cut}(\sigma) = V(\sigma), \quad \partial_{r}V_{cut}(\sigma)
  &= \partial_{r}V(\sigma), \quad \partial_{rr}V_{cut}(\sigma)
  = \partial_{rr}V(\sigma), \\
  V_{cut}(r_{c}) &= 0, \quad \partial_{r}V_{cut}(r_{c})=0.
\end{split}
\end{align}

In the definition of the cut-off potential (\ref{CutPot}), we only
consider even power polynomial terms of the atomic radius $r$ such
that we do not need to compute square roots of atomic distances in the
simulations.  The particular choice of conditions (\ref{Cond}) implies
that the harmonic approximation of the cut-off potential
(\ref{CutPot}) is equal to the harmonic potential approximation of
$V(r)$:
\[
V_{h}(r) = -\epsilon + 36\epsilon \left(\frac{r}{\sigma}-1\right)^2.
\]
Thus the linear analysis of the system for the nearest neighbour
interactions with potential $V_{cut}(r)$ is equivalent to the linear
analysis of the system with the original potential $V(r)$.

\begin{figure}[ht]
\centering 
\subfigure[]{\label{fig:InteractA} \includegraphics[width=0.48\textwidth]
{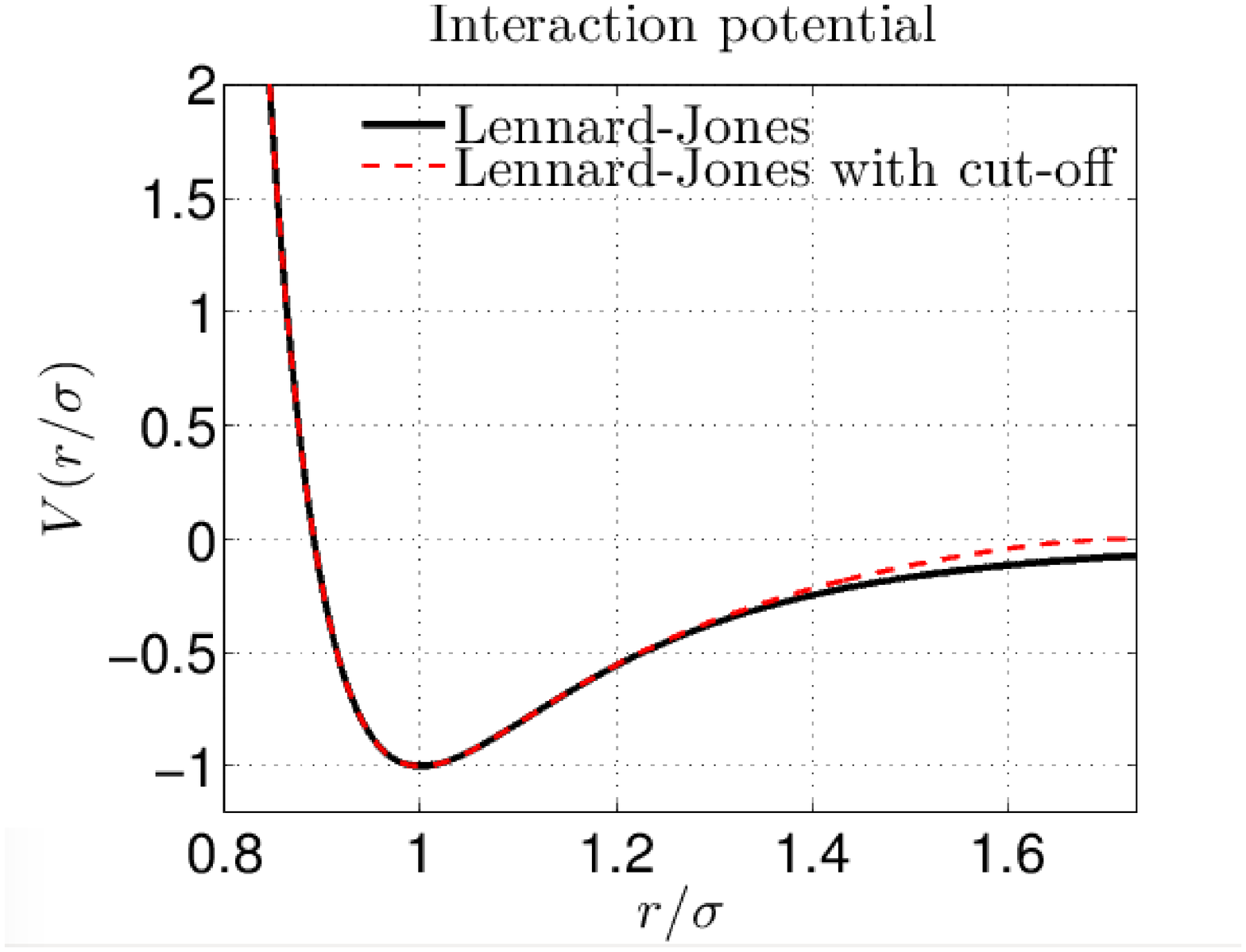}}
\subfigure[]{\label{fig:InteractB} \includegraphics[width=0.48\textwidth]
{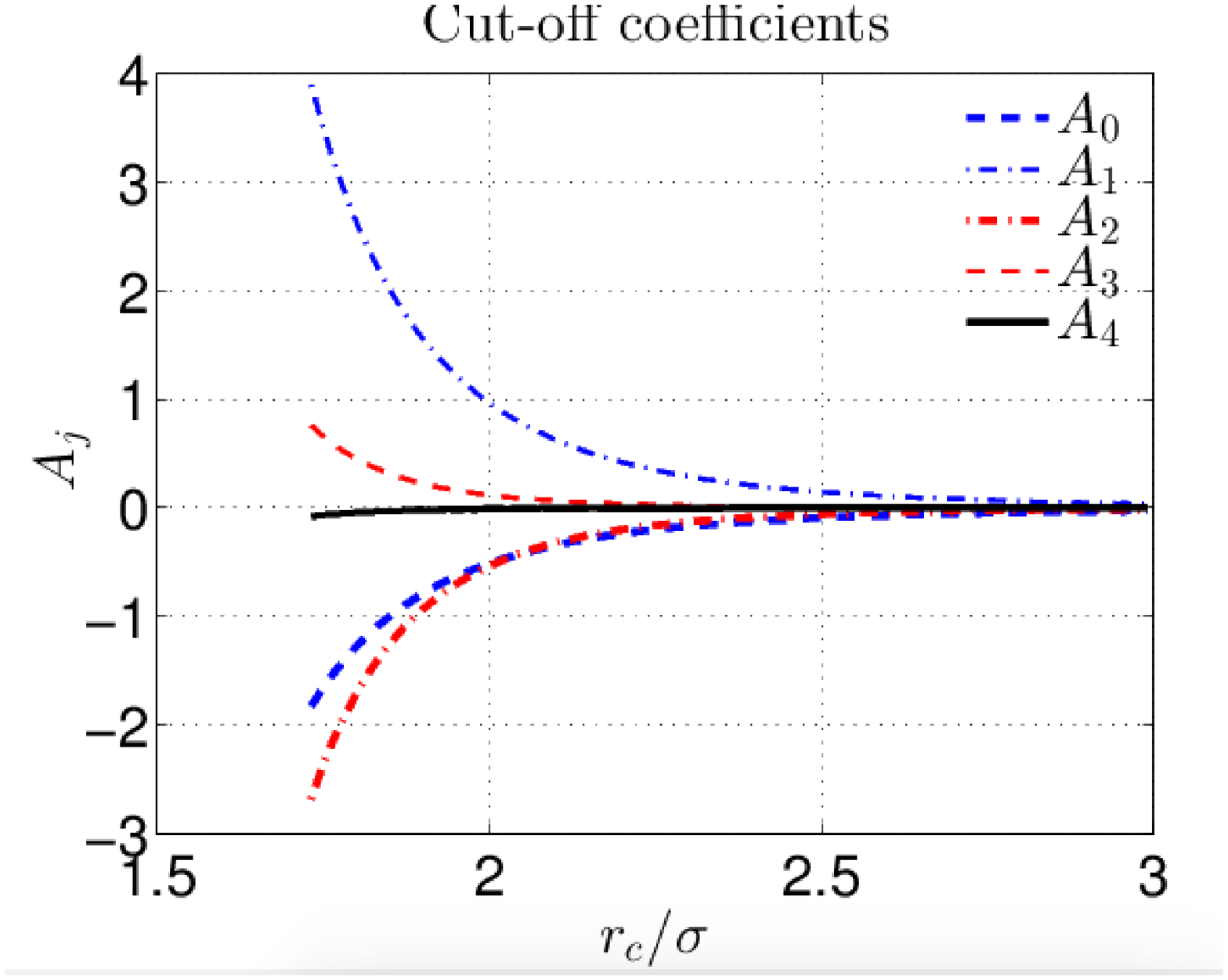}} 
\caption{Radial interaction potential $V(r)$. (a) Lennard-Jones
  potential compared to the Lennard-Jones potential with cut-off
  radius $r_{c}=\sqrt{3}\sigma$. (b) cut-off coefficients for the
  Lennard-Jones potential as functions of the cut-off radius
  $r_{c}/\sigma$.}\label{fig:Interact}
\end{figure}

In \ref{App:A}, we give exact formulas for the cut-off coefficients
$A_{j}$ of an arbitrary potential $V(r)$ satisfying the following
properties: $V\to{0}$ and $r\partial_{r}V\to{0}$ when $r\to\infty$.
The Lennard-Jones potential (\ref{LJpot}) satisfies these two
properties.  In Figure \ref{fig:InteractA} we compare the
Lennard-Jones potential (\ref{LJpot}) with the Lennard-Jones potential
with cut-off radius $r_{c}=\sqrt{3}\sigma$ computed by (\ref{CutPot}).
Due to the construction, the potential well is very well preserved
despite the additional polynomial terms in the potential. With Figure
\ref{fig:InteractB} we confirm that the cut-off coefficients $A_{j}$
tend to zero when the cut-off radius $r_{c}$ tends to infinity. Thus
in the limit we recover the original Lennard-Jones potential
(\ref{LJpot}).

\section{Dimensionless system of equations}
\label{sec:DimSyst}
In this section we derive a dimensionless system of equations.  We
consider the Hamiltonian (\ref{Hamilt}) with the on-site potential
(\ref{OnSiteFunc}) and the interaction potential (\ref{CutPot}), that
is, a system with total energy
\begin{align*}
H = & \sum_{n=1}^{N} \biggl( \frac{1}{2} m ||\dot{\B{r}}_{n}||^2 
+ U(U_{0},\sigma,\B{r}_{n}) \biggr. \\
 & \quad + \frac{1}{2}  \sum_{n'=1,\,n'\neq{n}}^{N} \biggl. 
\biggl( V(\epsilon,\sigma,r_{n,n'}) 
+ \epsilon \sum_{j=0}^{4} A_{j} \biggl( \frac{r_{n,n'}}{r_{c}} 
\biggr)^{2j} \biggr) \biggr),
\end{align*}
where $r_{n,n'}=||\B{r}_{n}-\B{r}_{n'}||$ and the potentials are
represented with their set of parameters and variables.  We introduce
a characteristic length scale $\sigma$ and time scale $T$ of the
system, i.e.~$\B{r}_{n}=\sigma\bar{\B{r}}_{n}$ and $t=T\bar{t}$.  Thus
$r_{n,n'}=\sigma\bar{r}_{n,n'}$, $r_{c}=\sigma\bar{r}_{c}$,
$\dot{\B{r}}_{n}=\sigma/T\dot{\bar{\B{r}}}_{n}$ and
$H=m\sigma^2/T^2\bar{H}$, where $\bar{H}$ is the dimensionless
Hamiltonian function.  Choosing the time scale $T=\sigma\sqrt{m/U_{0}}$
such that $H=U_{0}\bar{H}$, the dimensionless Hamiltonian $\bar{H}$ of
the dimensionless variables is
\begin{align*}
\bar{H} = & \sum_{n=1}^{N} \biggl( \frac{1}{2}
  ||\dot{\bar{\B{r}}}_{n}||^2 
+ U\left(1,1,\bar{\B{r}}_{n}\right)  \biggr.\\
 & \quad+ \frac{1}{2} \biggl. \sum_{n'=1,\,n'\neq{n}}^{N} 
\biggl( V\left(\bar{\epsilon},1,\bar{r}_{n,n'}\right) 
+ \bar{\epsilon} \sum_{j=0}^{4} A_{j} \left( \frac{\bar{r}_{n,n'}}
{\bar{r}_{c}} \right)^{2j} \biggr) \biggr),
\end{align*}
where $\bar{\epsilon}=\epsilon/U_{0}$ is a dimensionless parameter,
the interaction potential well depth parameter divided by the depth of
the on-site potential.  Dropping the bars over the variables, except
for the parameter $\bar{\epsilon}$, the dimensionless dynamical system
of equations is
\begin{align}\label{Systm}
\begin{split}
\dot{\B{r}}_{n} = & \B{u}_{n}, \\
\dot{\B{u}}_{n} = & -\partial_{\B{r}_{n}} U\left(1,1,\B{r}_{n}\right) \\
& - \frac{1}{2}\partial_{\B{r}_{n}} \sum_{n'=1,\,n'\neq{n}}^{N} 
\biggl( V\left(\bar{\epsilon},1,r_{n,n'}\right) 
+ \bar{\epsilon} \sum_{j=0}^{4} A_{j} \left( \frac{r_{n,n'}}
{r_{c}} \right)^{2j} \biggr) ,
\end{split}
\end{align}
for all $n=1,\dots,N$, where $\B{u}_{n} = \dot{\B{r}}_{n}$ is the
momentum. In the following we consider the dimensionless system
(\ref{Systm}) in our analysis and computations.

\begin{figure}[ht]
\centering 
\includegraphics[scale=0.42]{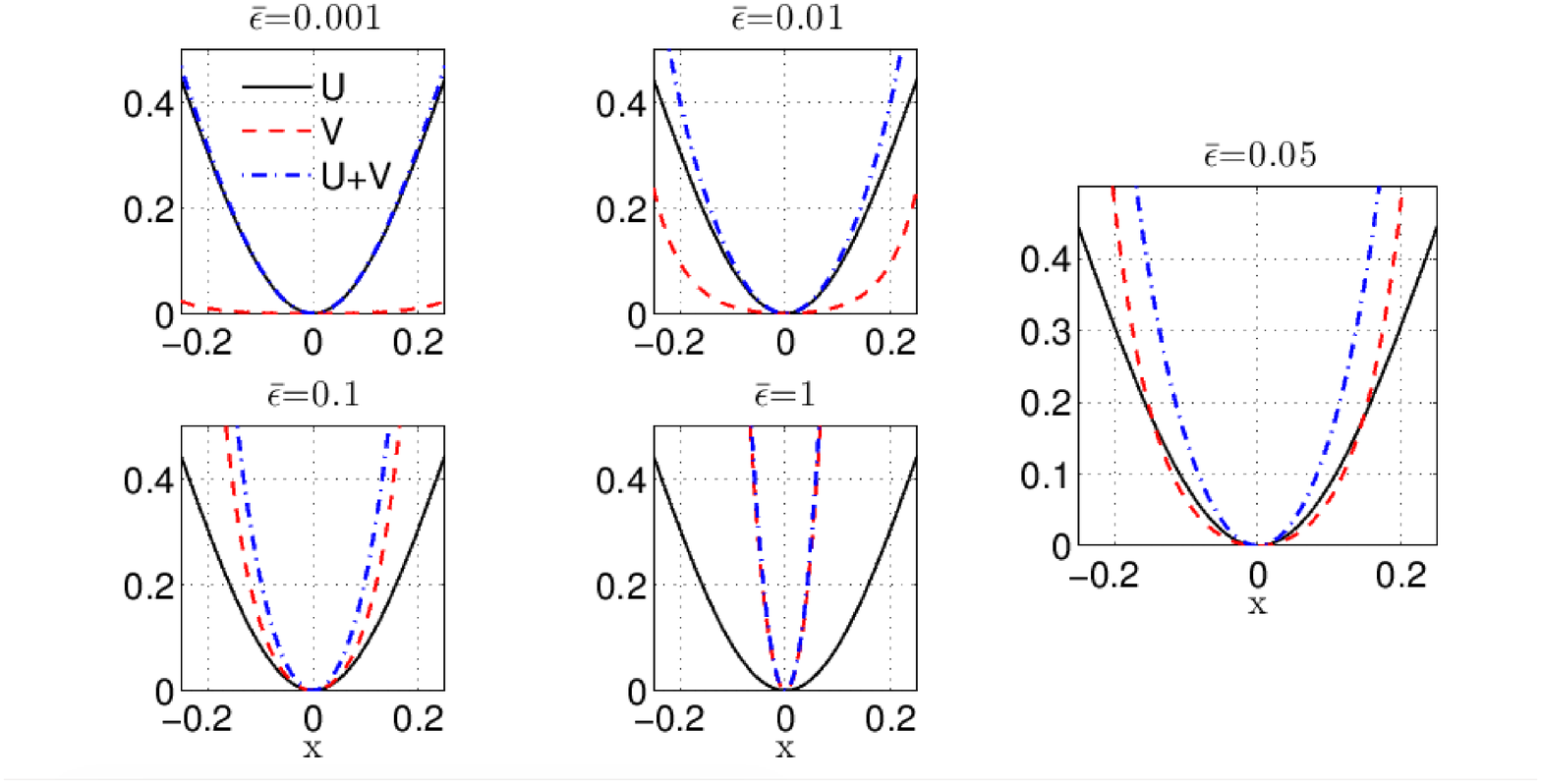}
\caption{Unrelaxed potential dependent on dimensionless parameter
  $\bar{\epsilon}$ as seen by a K atom moving in any of three
  lattice directions, illustrated in $(x,0)$ coordinates,
  $r_{c}=\sqrt{3}$.  The solid line shows the on-site potential $U$,
  the dashed line shows the Lennard-Jones interaction potential $V$
  normalized to positive values, and the dashed dotted line shows the
  sum of both potentials $U+V$.}\label{fig:UnrelaxedPot}
\end{figure}

The dimensionless system of equations (\ref{Systm}) contains two
dimensionless parameters $\bar{\epsilon}$ and $r_{c}$, and cut-off
coefficients $A_{j}$ which depend on cut-off radius $r_{c}$.
Independently from the value of $r_{c}$, when $\bar{\epsilon}=0$ there
is no interaction forces between potassium atoms and the system
(\ref{Systm}) decouples into a system of nonlinear oscillators.  When
the value of $\bar{\epsilon}$ tends to infinity, interaction forces
dominate over the forces from the on-site potential, and in this case
the equations describe a Lennard-Jones fluid.  To find a suitable
range for parameter $\bar{\epsilon}$ values such that both potentials
have relatively equal strength, we compute and compare unrelaxed
potentials seen by a potassium atom moving in any of the three lattice
directions.  In other words we fix all neighbouring atoms of a
particular K atom and compute potential energies for small atomic
displacements of the atom in any of three lattice directions, see
Fig.\ \ref{fig:UnrelaxedPot}.

In Figure \ref{fig:UnrelaxedPot} we show results for five parameter
$\bar{\epsilon}$ values.  It is evident that for $\bar{\epsilon} > 1$,
the interaction forces dominate the on-site forces, and for
$\bar{\epsilon}<0.001$ the interaction forces are too small compared
to the on-site forces and are negligible.  Assuming atomic relative
displacements from equilibrium in the range of $0.2$, these results
suggest that for system (\ref{Systm}) to model the K-K sheet
of muscovite mica we should choose $\bar{\epsilon}\in[1;0.001]$. We
find Fig.\ \ref{fig:UnrelaxedPot} to be in a good agreement with the
numerical results.  We do not observe propagating discrete breather
solutions outside of this range of $\bar{\epsilon}$ values.  Without loss
of generality we choose $\bar{\epsilon}=0.05$ as the main value for
our numerical studies.

\section{Linearised equations and dispersion relation}
\label{sec:LinEq}
In this section we derive a nearest neighbour elastic spring interaction
model of (\ref{Systm}) and its dispersion relation.  Recall that
the cut-off coefficients $A_{j}$ are chosen such that
$\partial_{rr}V_{cut}(\sigma)=\partial_{rr}V(\sigma)$, see conditions
(\ref{Cond}).  We consider the atom $\B{r}_{n}$ with labels $n=(l,m)$, see
Fig.\ \ref{fig:OnSiteB}, and its six neighbouring atoms with labels
$(l\pm{1},m)$, $(l+1,m\pm{1})$ and $(l-1,m\pm{1})$.  The force acting
on atom $n$ from atom $n'$ is given by a vector
\begin{equation}\label{Force}
 \B{F}_{n,n'} = -\frac{1}{r}\partial_{r}V(r) \left( \B{r}_{n} 
- \B{r}_{n'} \right),
\end{equation}
where $r\equiv r_{n,n'}=||\B{r}_{n} - \B{r}_{n'}||$. The linearised
version of (\ref{Force}) around the dynamical equilibrium states
$\B{r}_{n}^{0}$ and $\B{r}_{n'}^{0}$ with $r^{0}=||\B{r}_{n}^{0} -
\B{r}_{n'}^{0}||$ is
\[
\B{F}_{n,n'}^{lin} = \partial_{\B{r}_{n}}\B{F}_{n,n'} \left( \B{r}_{n}
  - \B{r}_{n}^{0} \right) + \partial_{\B{r}_{n'}}\B{F}_{n,n'}\left(
  \B{r}_{n'} - \B{r}_{n'}^{0} \right),
\]
where $\partial_{\B{r}_{n'}}\B{F}_{n,n'} =
-\partial_{\B{r}_{n}}\B{F}_{n,n'}$, and
\begin{align*}
  \partial_{\B{r}_{n}}\B{F}_{n,n'} &=
  - \partial_{rr}V\left(r^{0}\right)
  \frac{\B{r}_{n}^{0}-\B{r}_{n'}^{0}}{r^{0}}
  \left(    \frac{\B{r}_{n}^{0}-\B{r}_{n'}^{0}}{r^{0}} \right)^{T} \\
  & = - \partial_{rr}V\left(r^{0}\right) D_{n,n'} = -V_{r^{0}}^{''}
  D_{n,n'}.
\end{align*}
The vector $\left( \B{r}_{n}^{0} - \B{r}_{n'}^{0} \right) / r^{0}$ is
the corresponding direction cosine vector for the atomic pair $(n,n')$
equilibrium positions, six in this case, and the symmetric matrix
$D_{n,n'}\in\R^{2\times{2}}$ is an outer product of the direction
cosine vector.

Note that $V_{r^{0}}^{''}=72\bar{\epsilon}$ and $r^{0}=1$ are constant
when only nearest neighbour Lennard-Jones interactions are considered
and recall that the harmonic approximation of the on-site potential
(\ref{OnSiteFunc}) is given by (\ref{OnSiteFuncAppr}).  By dropping
index $n=(l,m)$ from matrix $D_{n,n'}$ and replacing index $n'$ with
the six neighbouring labels of atoms with labels $(l,m)$ from Fig.\
\ref{fig:OnSiteB}, we obtain a system of dynamical linear equations
\begin{align}\label{LinSyst}
\begin{split}
  \ddot{\B{w}}_{l,m} =& -\left( D_{l+2,m} + D_{l-2,m} + D_{l+1,m+1} \right.\\
  & \left.+ D_{l+1,m-1} + D_{l-1,m+1} + D_{l-1,m-1} \right) \B{w}_{l,m} \\
  & + D_{l+2,m} \B{w}_{l+2,m} + D_{l+1,m+1} \B{w}_{l+1,m+1}
  + D_{l-1,m+1} \B{w}_{l-1,m+1}\\
  & + D_{l-2,m} \B{w}_{l-2,m} + D_{l+1,m-1} \B{w}_{l+1,m-1}
  + D_{l-1,m-1} \B{w}_{l-1,m-1}\\
  & - \kappa \B{w}_{l,m},
\end{split}
\end{align}
where $\B{w}_{l,m} = \B{r}_{l,m}-\B{r}_{l,m}^{0}$ is the displacement
vector from the equilibrium state of atom $(l,m)$ and
$\kappa = 16\pi^2/9/V_{r^{0}}^{''}$ after time rescaling
$t=\hat{t}\sqrt{V_{r^{0}}^{''}}$.  In \ref{App:B} we give exact
expressions for matrices $D_{l,m}$ and system (\ref{LinSyst}) in
component-wise form, which are in exact agreement with the
linearised equations of the Morse hexagonal lattice with an on-site
harmonic potential presented in \cite{IkDoFeKa07}.

We argue here that the equation (\ref{LinSyst}) with linear
interaction forces together with the egg-box on-site potential
(\ref{OnSiteFunc}), instead of a harmonic on-site potential
(\ref{OnSiteFuncAppr}) is a 2D model with lower level of complexity
and could be used as a starting point for analytical investigations.
In particular, it may be possible to develop an existence proof for
propagating localized modes in this model.  The proposed model can be
thought as a natural extension of the discrete sine-Gordon equation
in two dimensions with hexagonal symmetry.  Similarly a square
lattice could be considered.

Following the approach of \cite{IkDoFeKa07}, we derive a
dispersion relation from the simple wave solutions
\begin{equation}\label{SimplWave}
 \B{w}_{l,m} = \B{A} e^{i\left( \frac{1}{2}k_{1}l +
     \frac{\sqrt{3}}{2}k_{2}m 
- \hat{\omega}\hat{t}\right)},
\end{equation}
where $\B{A}\in\R^{2}$ is an amplitude, $\B{k}=(k_{1},k_{2})$ is the
wave number, and $\hat{\omega}$ is a frequency in the $\hat{t}$ time
scale.  Explicit calculations give the linear system matrix for
$\B{w}_{l,m}$ in (\ref{LinSyst})
\begin{align}\label{SystMatrix}
\begin{split}
D_{l+2,m} &+ D_{l-2,m} + D_{l+1,m+1} \\
&+ D_{l+1,m-1} + D_{l-1,m+1} + D_{l-1,m-1} = 
\begin{pmatrix} 3 & 0 \\ 0 & 3 \end{pmatrix},
\end{split}
\end{align}
see \ref{App:B}. Thus, substituting (\ref{SimplWave}) and
(\ref{SystMatrix}) into the linear system (\ref{LinSyst}), we obtain
\begin{align}\label{EqA}
\begin{split}
\B{0} = & \left(\hat{\omega}^2 - \kappa - 3 \right)\B{A} 
+ D_{l+2,m} \B{A} e^{+ik_{1}} + D_{l-2,m} \B{A} e^{-ik_{1}}\\
& + D_{l+1,m+1} \B{A} e^{i\left(  \frac{1}{2}k_{1} + \frac{\sqrt{3}}{2}k_{2} \right)} 
  + D_{l+1,m-1} \B{A} e^{i\left(  \frac{1}{2}k_{1} - \frac{\sqrt{3}}{2}k_{2} \right)}\\
& + D_{l-1,m+1} \B{A} e^{i\left( -\frac{1}{2}k_{1} + \frac{\sqrt{3}}{2}k_{2} \right)} 
  + D_{l-1,m-1} \B{A} e^{i\left( -\frac{1}{2}k_{1} - \frac{\sqrt{3}}{2}k_{2} \right)}.
\end{split}
\end{align}
From the symmetry properties of matrices $D_{l,m}$, equation
(\ref{EqA}) can be simplified to
\begin{align*}
\begin{split}
  \B{0} =& \left(\hat{\omega}^2 - \kappa - 3 \right)\B{A} \\
  & + 2\cos\left(k_{1}\right)\begin{pmatrix} 1 & 0\\0 &
    0 \end{pmatrix} \B{A} + \cos \left( \tfrac{1}{2}k_{1} \right) \cos
  \left( \tfrac{\sqrt{3}}{2}k_{2} \right)
\begin{pmatrix} 1 & 0\\0 & 3 \end{pmatrix}  \B{A} \\
&- \sin \left( \tfrac{1}{2}k_{1} \right) \sin 
\left( \tfrac{\sqrt{3}}{2}k_{2} \right)
\begin{pmatrix} 0 & \sqrt{3} \\ \sqrt{3} & 0 \end{pmatrix} \B{A}
\end{split}
\end{align*}
and this leads to the dispersion relation
\begin{align}\label{DispRel}
\begin{split}
  \left(\hat{\omega}^2 - \kappa - 3 + 2\cos\left(k_{1}\right) + \cos
    \left( \tfrac{1}{2}k_{1} \right) \cos
    \left( \tfrac{\sqrt{3}}{2}k_{2} \right) \right) &\\
  \times \left(\hat{\omega}^2 - \kappa - 3 + 3\cos \left(
      \tfrac{1}{2}k_{1} \right) \cos
    \left( \tfrac{\sqrt{3}}{2}k_{2} \right)  \right) &\\
  -3\sin^2 \left( \tfrac{1}{2}k_{1} \right) \sin^2 \left(
    \tfrac{\sqrt{3}}{2}k_{2} \right) &= 0,
\end{split}
\end{align}
which is in the exact agreement with the dispersion relation obtained
for the Morse hexagonal lattice in \cite{IkDoFeKa07}.

\begin{figure}[ht]
\centering 
\subfigure[]{\label{fig:DispRelA}\includegraphics[width=0.48\textwidth]
{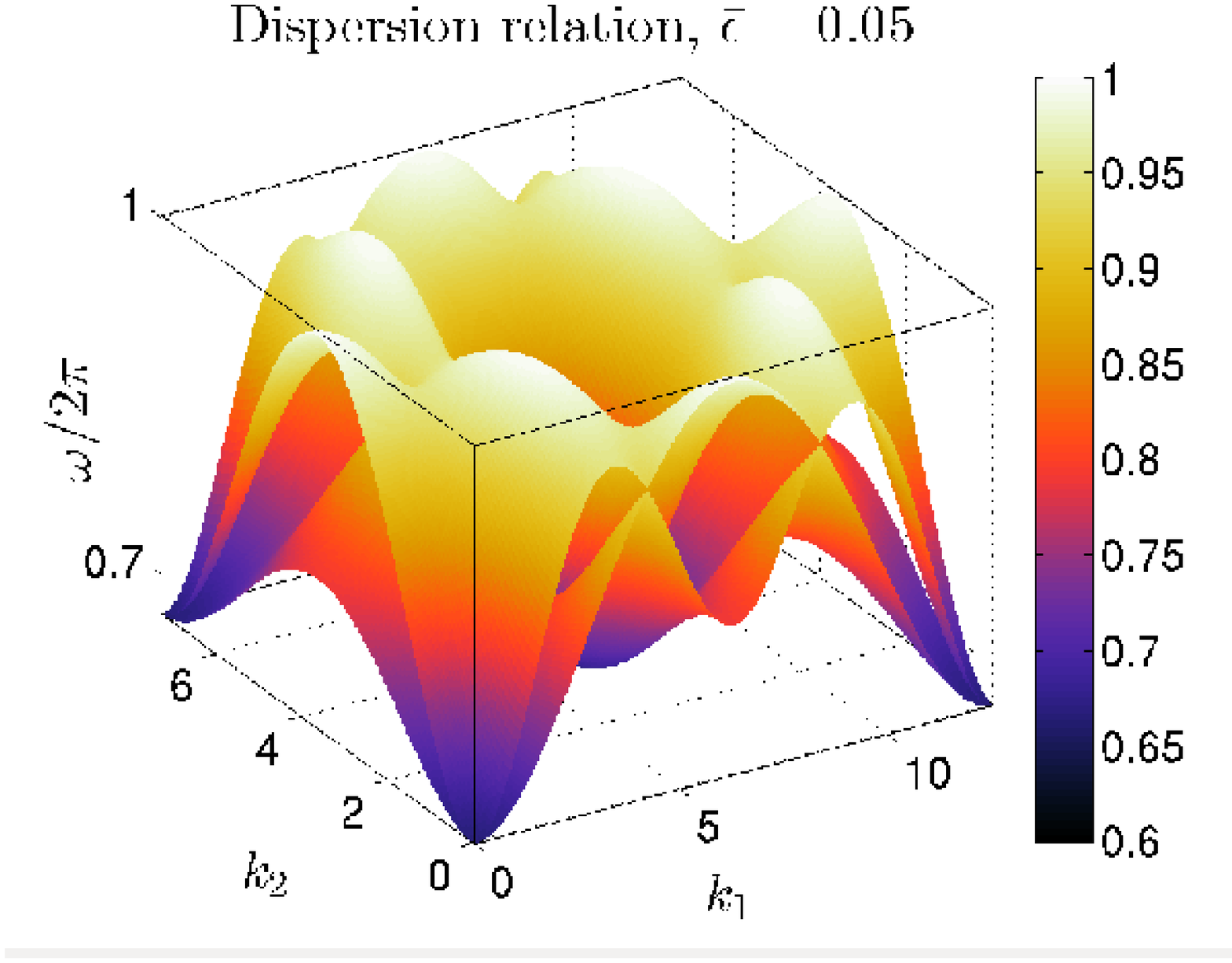}}
\subfigure[]{\label{fig:DispRelB}\includegraphics[width=0.48\textwidth]
{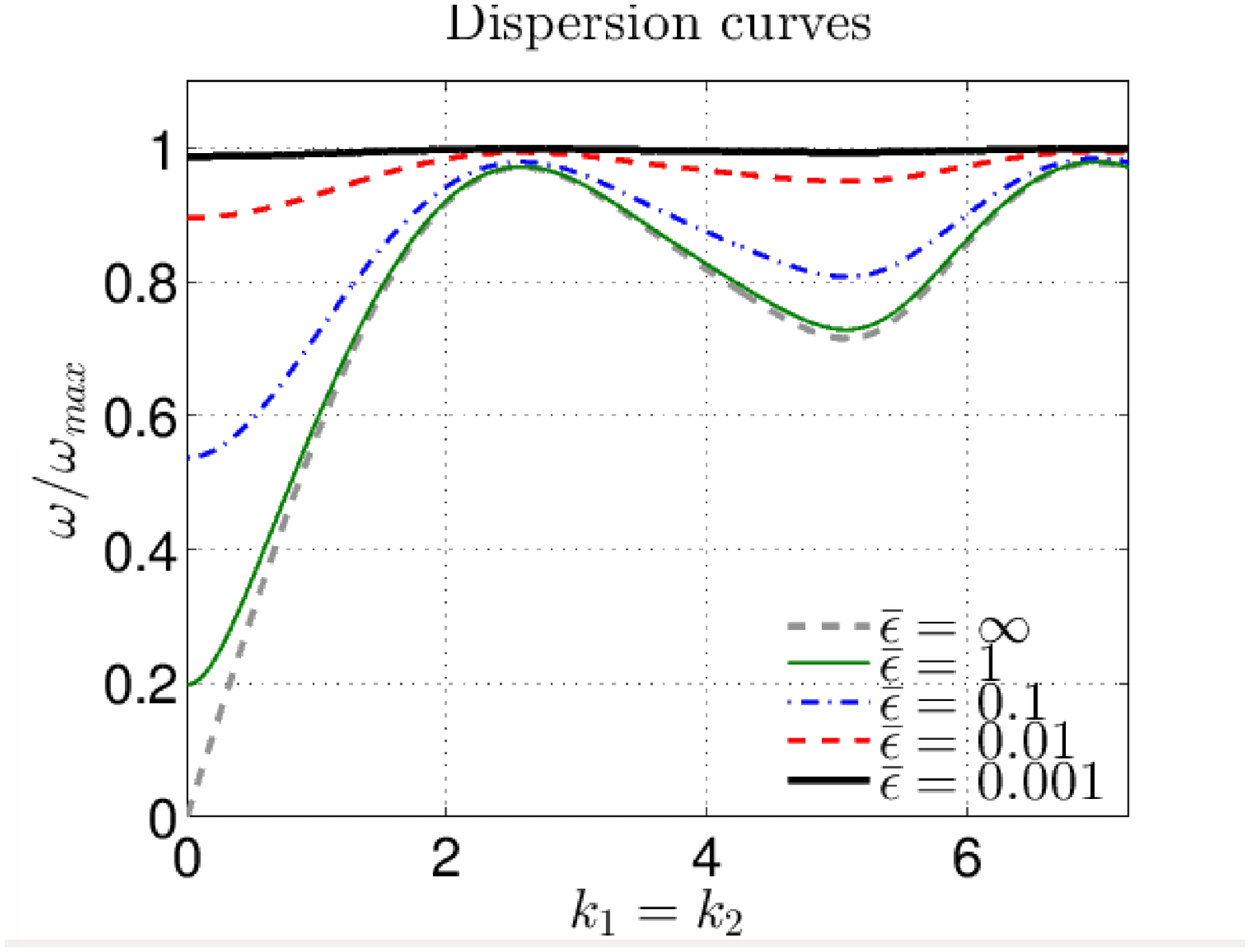}} 
\caption{Dispersion relation of the linearised 2D hexagonal crystal
  lattice equations. (a) two branches of positive frequency
  $\omega/2\pi$, $\bar{\epsilon}=0.05$. (b) normalized upper branch
  dispersion curves for equal components of the wavenumber,
  i.e.~$k_{1}=k_{2}$, and different values of
  $\bar{\epsilon}$.}\label{fig:DispRel}
\end{figure}

The frequency solution of (\ref{DispRel}) has two positive branches
for a given wavenumber $\B{k}=(k_{1},k_{2})$.  From (\ref{EqA}), by
setting $\B{k}=(2\pi,0)$, for example, we can derive the maximal
frequency of the linear system (\ref{LinSyst}) in $\hat{t}$ and $t$
time scales, that is
\[
\hat{\omega}_{max} =\sqrt{6+\kappa}, \quad 
\omega_{max} =\sqrt{6V_{r^{0}}^{''}+\frac{16\pi^2}{9}}, \quad
\omega = \sqrt{V_{r^{0}}^{''}}\hat{\omega},
\]
respectively. In Figure \ref{fig:DispRelA} we show surface plots of
frequency $\omega/2\pi$ versus wavenumber for $\bar{\epsilon}=0.05$.
In Figure \ref{fig:DispRelB}, we plot normalized dispersion curves for
equal components of wavenumber, that is, $k_{1}=k_{2}$, for different
values of $\bar{\epsilon}$.  The normalized frequency can be
expressed as
\[
 \frac{\omega}{\omega_{max}} = \frac{\hat{\omega}}{\hat{\omega}_{max}} 
= \frac{\sqrt{\alpha(k_{1},k_{2}) + \kappa(\bar{\epsilon})}}
{\sqrt{6+\kappa(\bar{\epsilon})}} 
\]
which tends to unity when $\bar{\epsilon}\to{0}$ for each value of
$(k_{1},k_{2})\in\R^2$.  Hence $\omega\to\omega_{max}=4\pi/3$ when
$\bar{\epsilon}\to{0}$, the case of a decoupled system of harmonic
oscillators of potential energy (\ref{OnSiteFuncAppr}) with $U_{0}=1$.

\section{Numerical simulations of propagating 
discrete breathers}
\label{sec:NumSim}
In this section we describe numerical simulations of propagating
discrete breathers obtained by solving the initial value problem
(\ref{Systm}).  We integrate the Hamiltonian dynamics (\ref{Systm}) in
time with a second order time reversible symplectic Verlet method
\cite{AlTi89,LeRe05}.  In the following, all numerical examples are
performed with $\bar{\epsilon}=0.05$, $r_{c}=\sqrt{3}$, time step
$\tau=0.04$ and periodic boundary conditions. To excite mobile
discrete breathers, we consider the lattice in its dynamical
equilibrium state, see Fig.\ \ref{fig:OnSiteB}, and excite three
neighbouring atomic momenta with the pattern
\begin{equation}\label{Pattern}
 \B{v}_{0} = \gamma (-1;2;-1)^{T},
\end{equation}
where the values of $\gamma \ne 0$ depend on the choice of
$\bar{\epsilon}$.  Single kicks or different patterns of simultaneous
kicks can be considered as well as the excitation given above.  Our
objective is to consider initial conditions which produce the least
amount of phonon background which may interfere with the study of
propagating discrete breathers.  This particular choice of pattern
(\ref{Pattern}) gave the cleanest initial conditions for the
computations of propagating discrete breathers. The following results
are presented with $\gamma=0.5$.

The main observed properties of propagating discrete breathers were
qualitatively similar for different values of the cut-off radius
$r_{c}$, for $\bar{\epsilon}$ values when propagating discrete
breathers can be observed, time steps $\tau$, for different initial
momenta patterns and for different values of $\gamma$.  However, long
time numerical simulations are sensitive to initial conditions, small
changes in parameter and time step values as well as to round-off
errors. This is due to the chaotic nature of the underlying dynamical system.

To display the energy over the lattice, we define an energy density
function by assigning to each atom its kinetic energy and on-site
potential values as well as half of the interaction potential values.
Since the energy $H$ may take also negative values, and to explore
better small scales of the system for plotting purposes only, we
replace the total energy of the system by
\[
 H_{log} = \log(H + |\min{\{H\}}| + 1)
\]
such that $H_{log}\ge 0$.

\begin{figure}[ht]
\centering 
\includegraphics[width=0.32\textwidth]{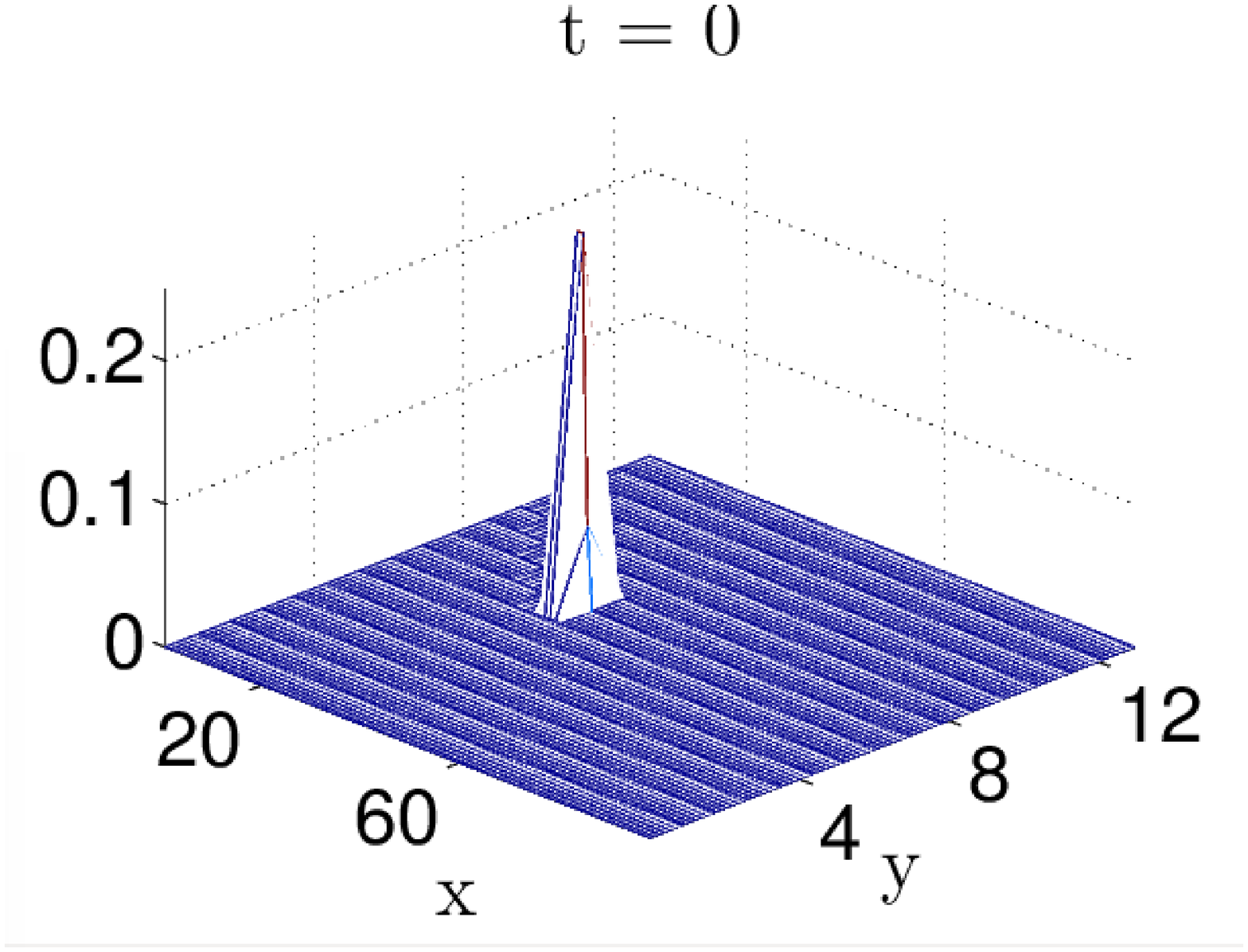}
\includegraphics[width=0.32\textwidth]{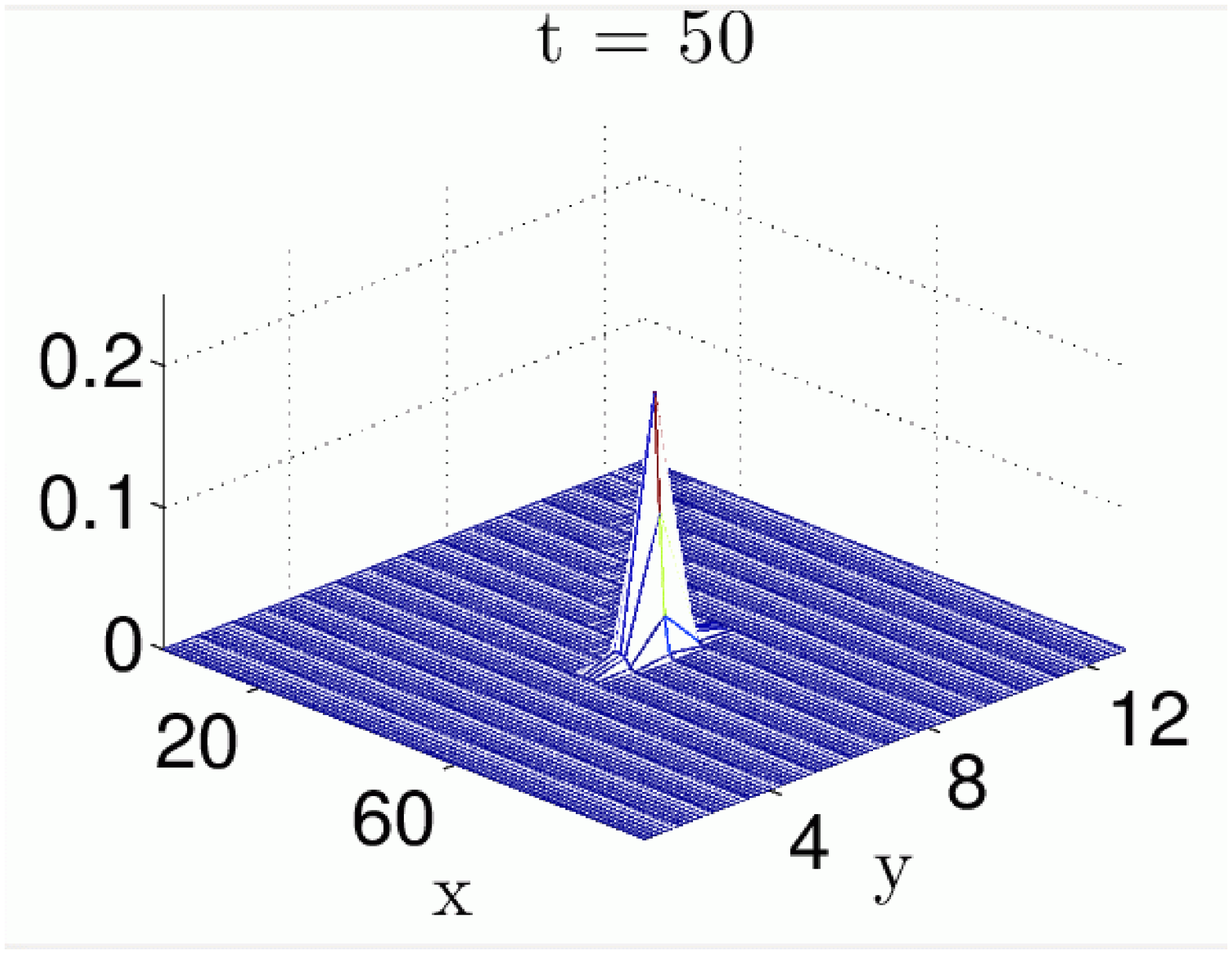} 
\includegraphics[width=0.32\textwidth]{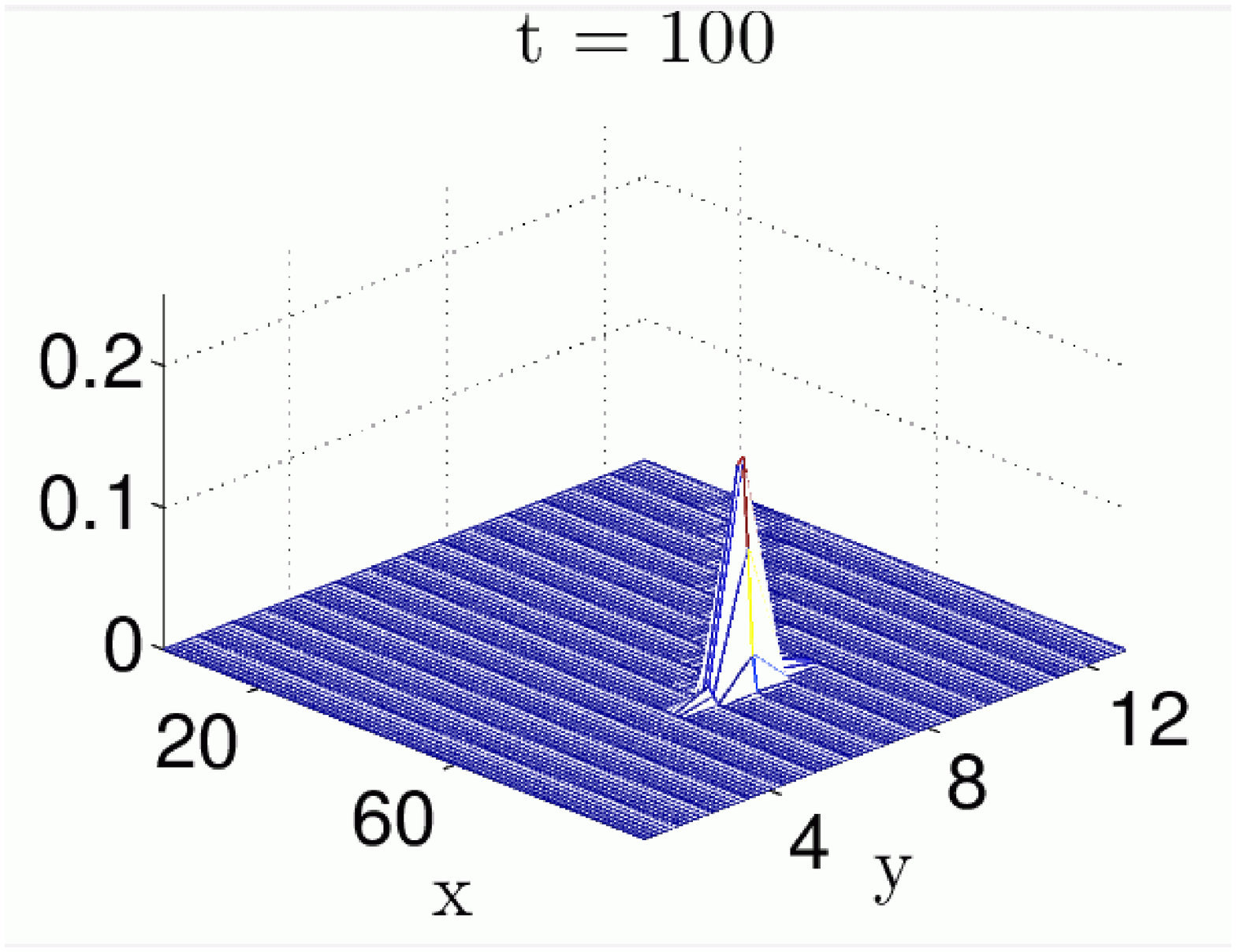} 
\includegraphics[width=0.32\textwidth]{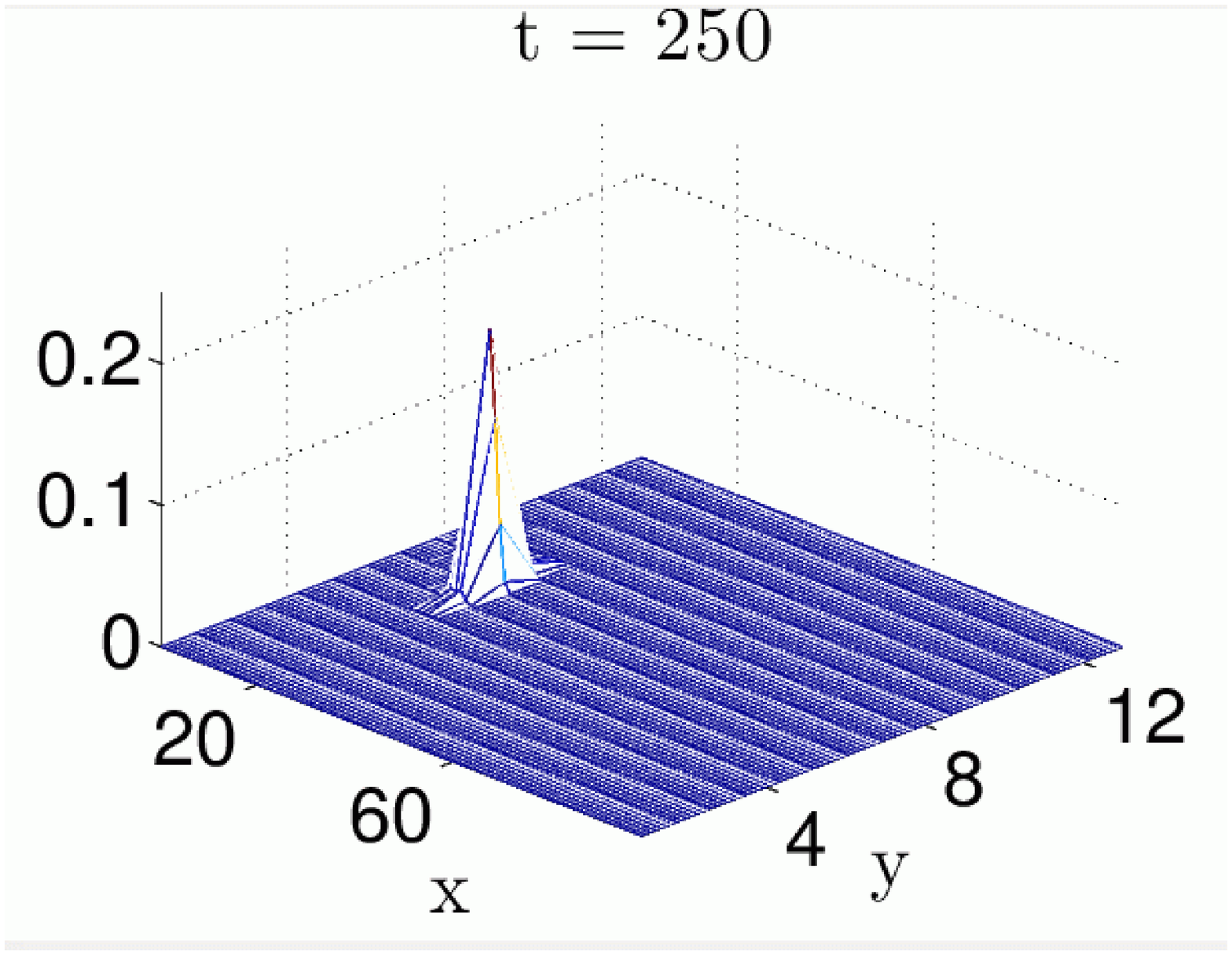}
\includegraphics[width=0.32\textwidth]{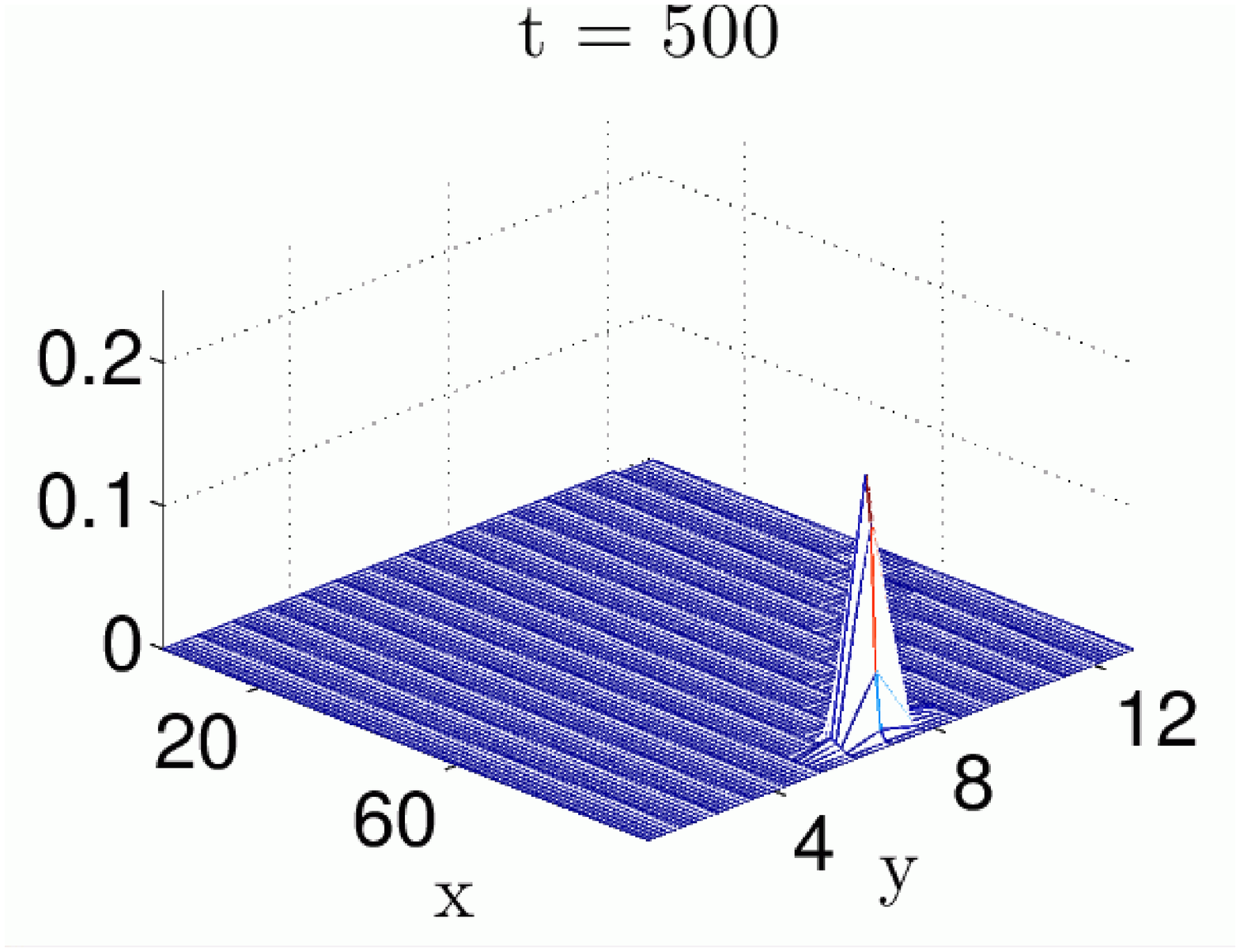} 
\includegraphics[width=0.32\textwidth]{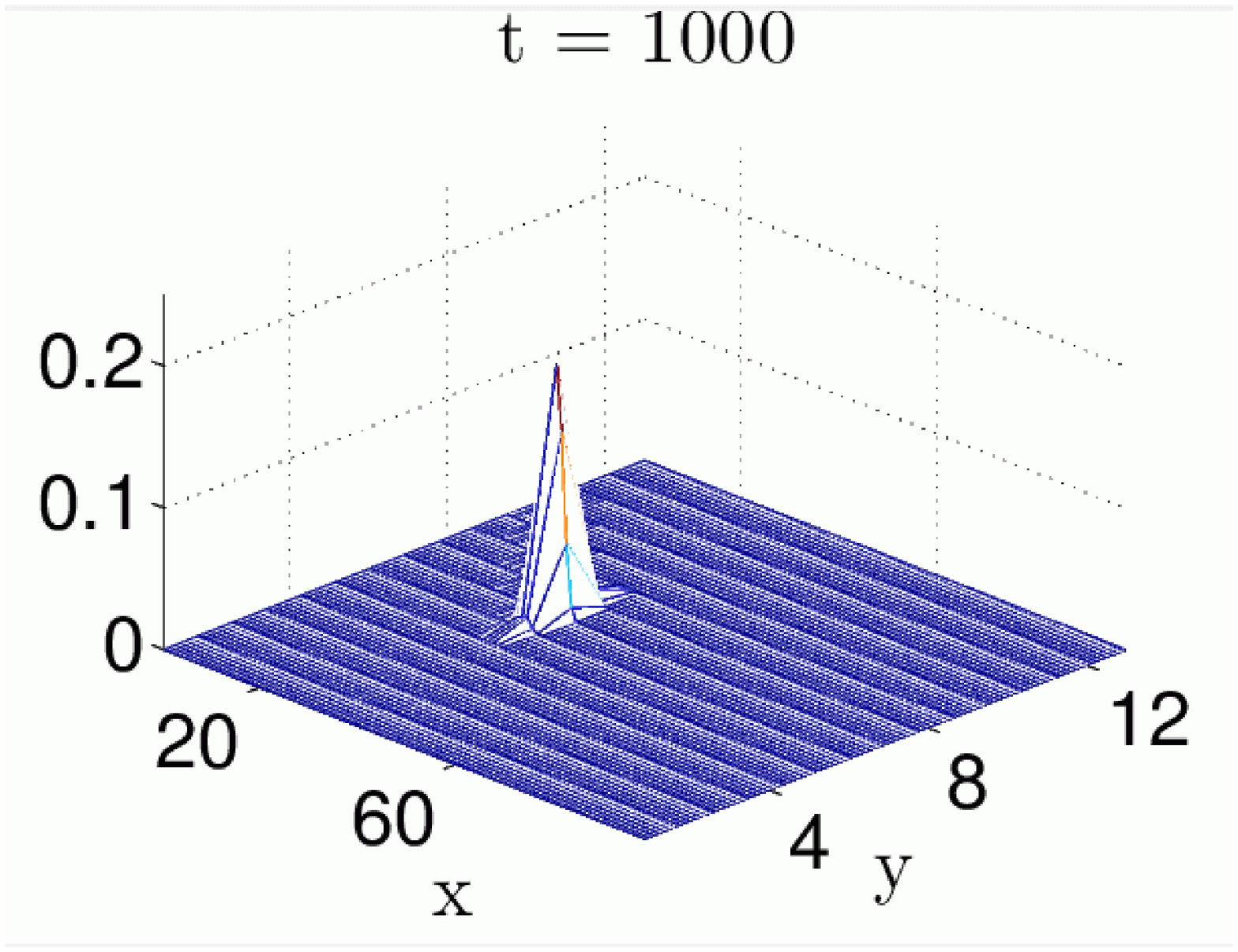} 
\caption{Evolution of the energy density function in time and energy
  localization around the propagating discrete breather
  solution.}\label{fig:3DEn}
\end{figure}

As a first example, we consider a periodic rectangular lattice:
$N_{x}=100$ and $N_{y}=16$, where $N_{x}$ and $N_{y}$ are the number
of atoms in $x$ and $y$ axis directions, respectively.  We place the
initial momentum pattern (\ref{Pattern}) in the middle of the domain
with respect to the $y$ axis.  We integrate the system in time up to
$1000$ time units.  In Fig.\ \ref{fig:3DEn} we demonstrate the
evolution of the energy density function in time.  For plotting
purposes we have interpolated the energy density function on a
rectangular mesh.  The peaks of energy in Fig.\ \ref{fig:3DEn} are
associated with the propagating discrete breather.  To perform this
test, we also included a damping of the atomic momenta at the upper and
lower boundaries for the initial time interval $t\in[0;100]$, to reduce
the amount of phonons which spread over the domain.  The propagating
breather moves to the right on the horizontal line, i.e.~on the
horizontal crystallographic lattice line, and is highly localized in
space.  Evidently, from Fig.\ \ref{fig:3DEn}, the initial pattern
(\ref{Pattern}) with a small amount of initial damping at the boundaries
for some time interval, has created a clean breather solution with
small amplitude phonon background not visible to the naked eye.

\begin{figure}[ht]
\centering 
\subfigure[]{\label{fig:BDisplA}\includegraphics[width=0.48\textwidth]
{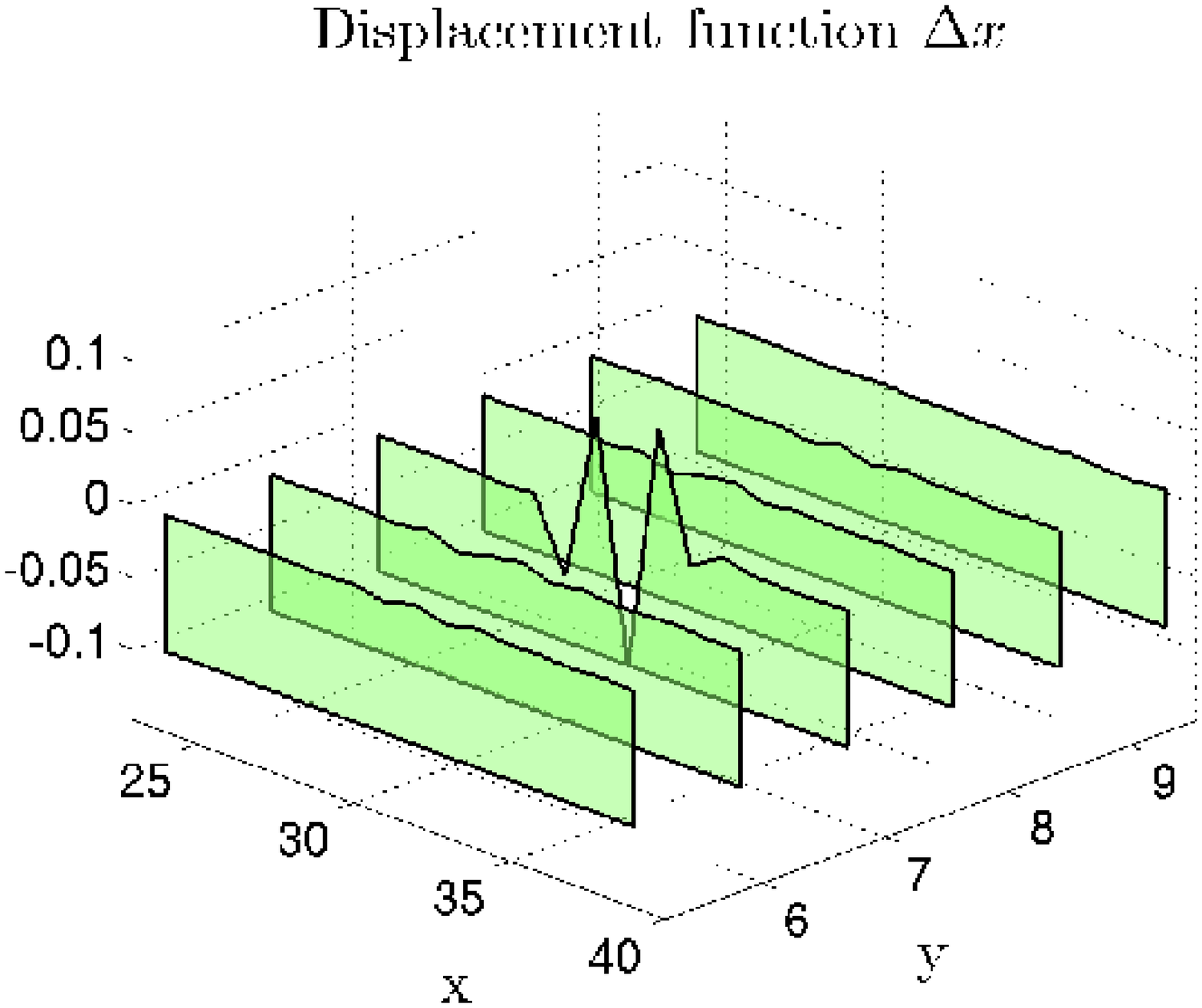}}
\subfigure[]{\label{fig:BDisplB}\includegraphics[width=0.48\textwidth]
{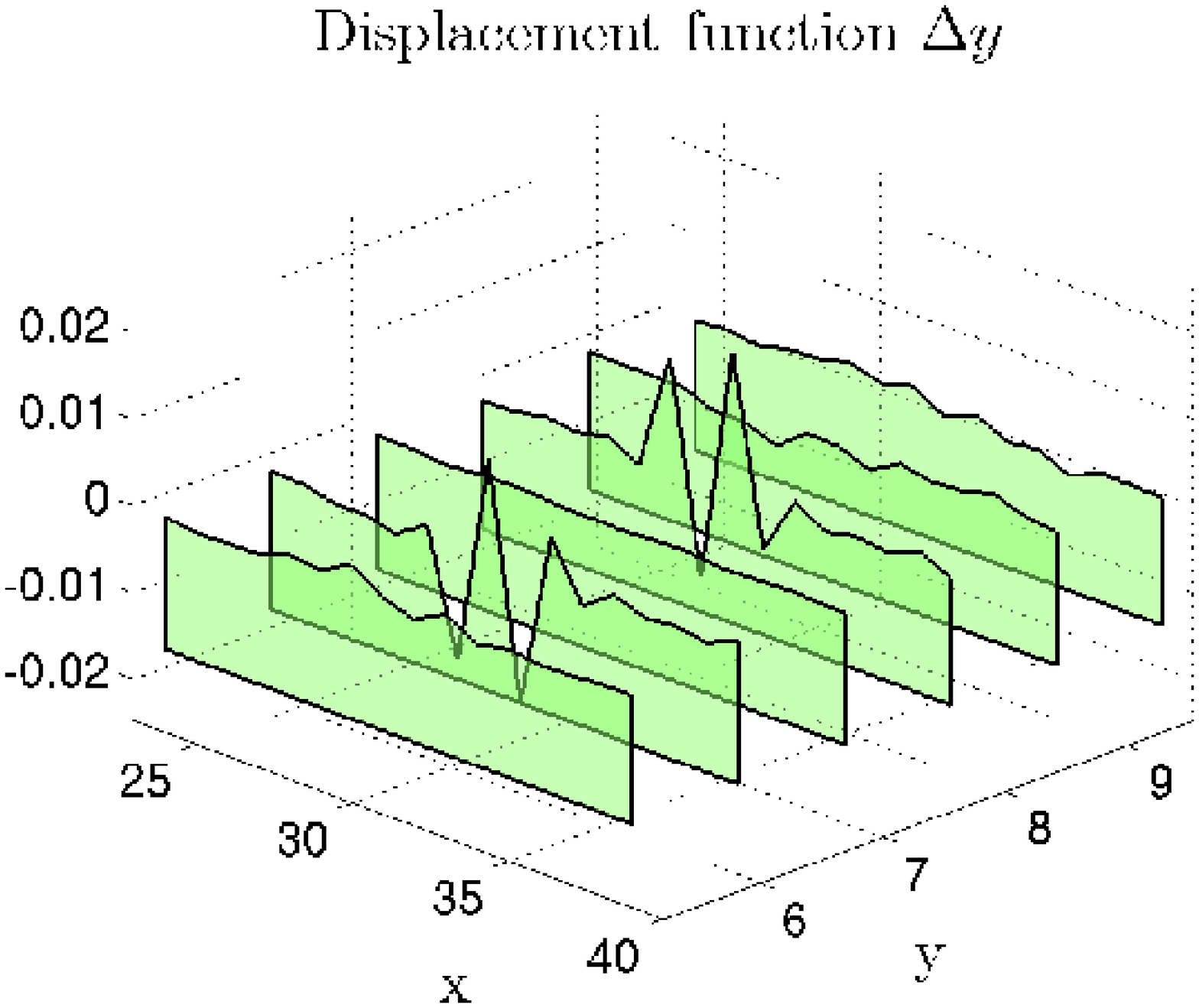}} 
\caption{Atomic displacements in space from their equilibrium states at
  final computational time $T_{end}=1000$. (a) displacement function
  $\Delta{x}$ in the $x$ axis direction. (b) displacement function
  $\Delta{y}$ in the $y$ axis direction.}\label{fig:BDispl}
\end{figure}

In Figure \ref{fig:BDispl} we plot atomic displacements from their
equilibrium states in the $x$ and $y$ axis directions at the final
computational time, i.e.~$T_{end}=1000$.  We indicate the displacement
function in the $x$ axis direction by $\Delta{x}$ and in the $y$ axis
direction by $\Delta{y}$.  Comparing Figures \ref{fig:BDisplA} and
\ref{fig:BDisplB}, we notice the differences in the scales of the
displacements.  From Figure \ref{fig:BDisplA}, it is evident that the
largest displacement in the $x$ direction is on the main chain of
atoms along which the breather propagates and there are only very
small amplitude displacements in adjacent chains.  From Figure
\ref{fig:BDisplB}, it can be seen that on the main chain there is
almost zero displacement in the $y$ direction, whilst there is visible
displacement of the adjacent chains.  Notice the anti-symmetry between
breather displacements on adjacent chains in Fig.\ \ref{fig:BDisplB}.

\begin{figure}[ht]
\centering 
\subfigure[]{\label{fig:DisplA}\includegraphics[width=0.48\textwidth]
{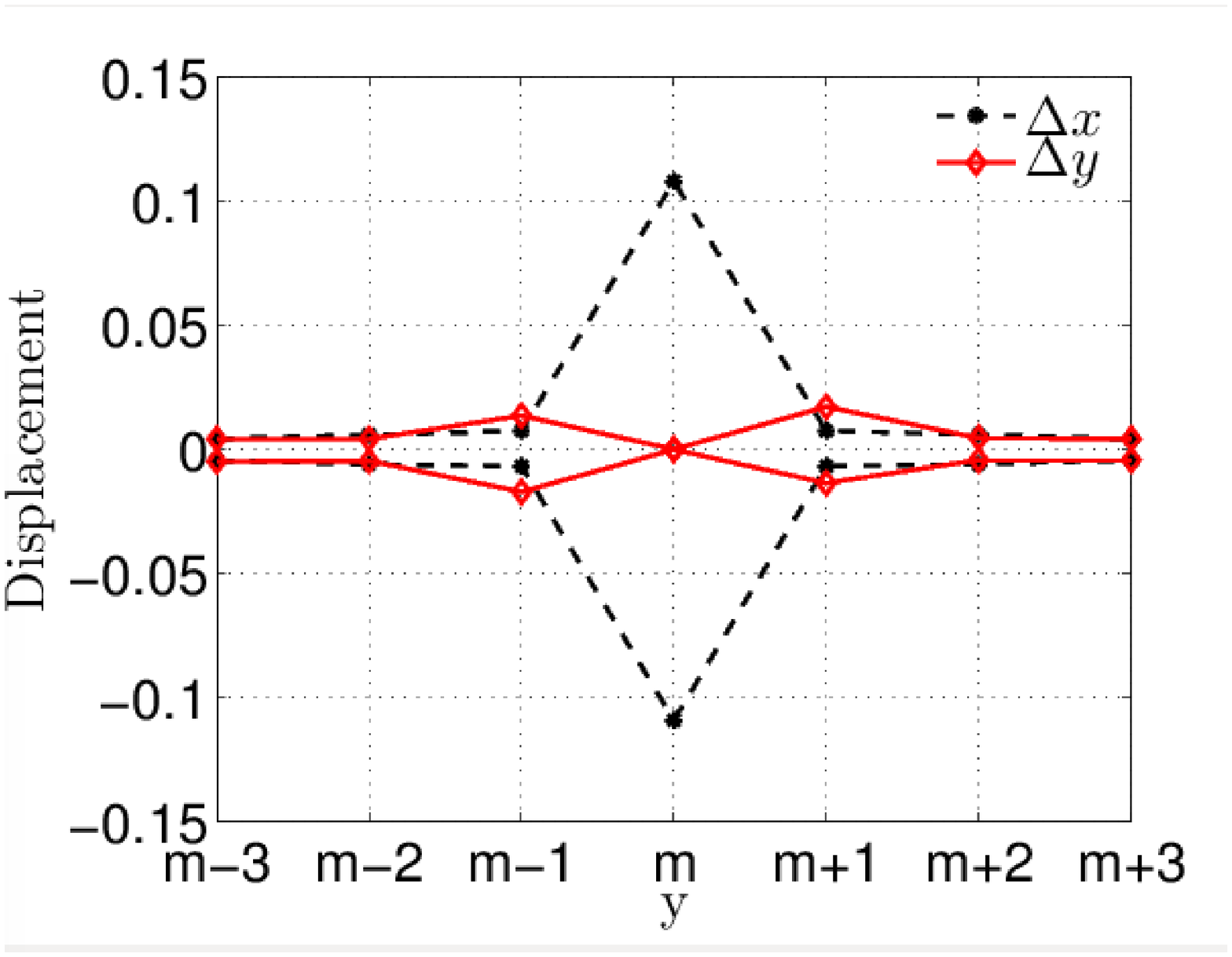}}
\subfigure[]{\label{fig:DisplB}\includegraphics[width=0.48\textwidth]
{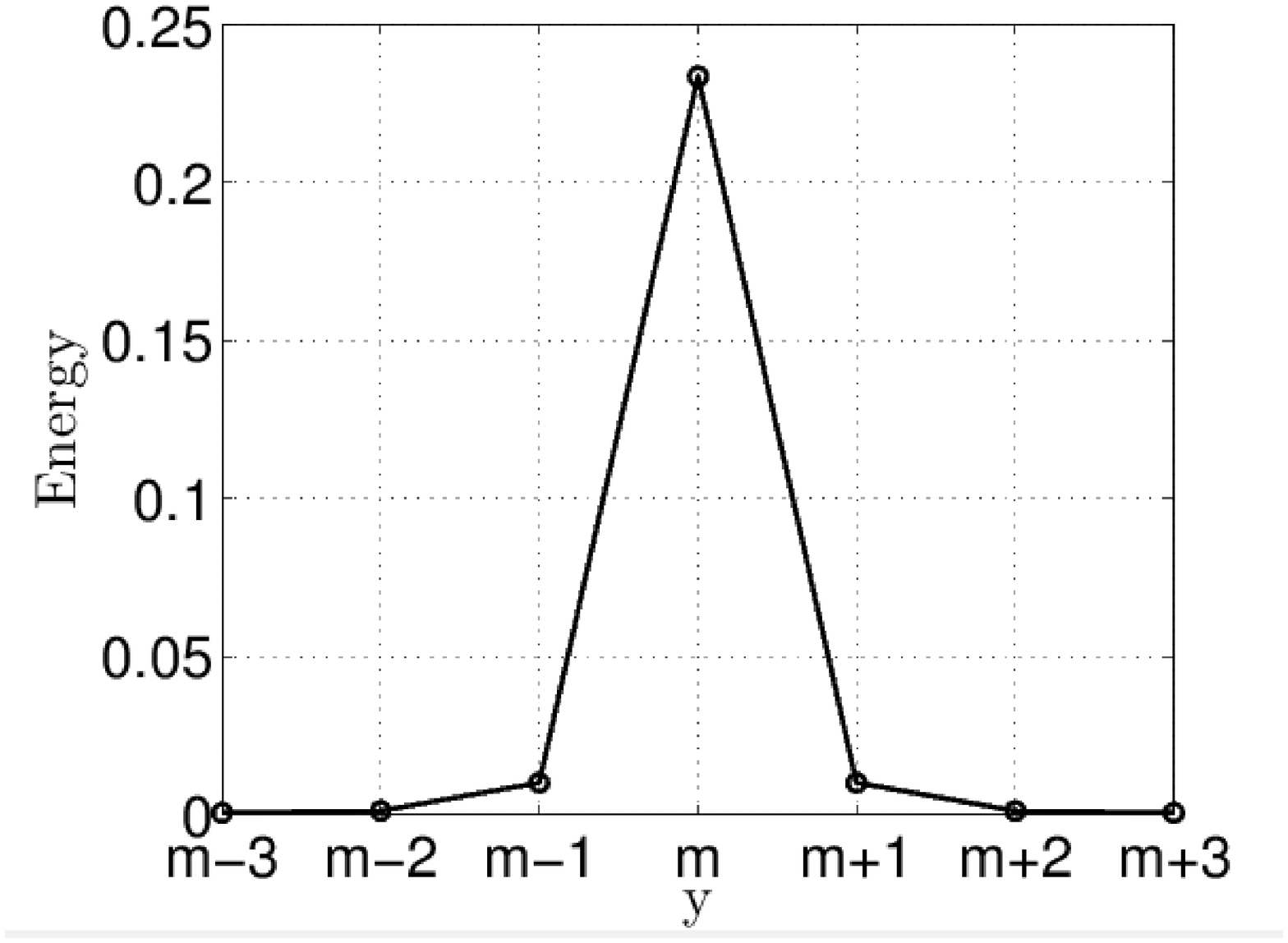}} 
\caption{Spacial displacements and energy of the propagating discrete
  breather. (a) maximal and minimal displacements in $x$ and $y$ axis
  directions on atomic chains over computational time interval. (b)
  maximal energy in chains of atoms over computational time
  interval.}
\label{fig:Displ}
\end{figure}

To understand better the localization properties of the propagating
breather solutions, we compute the maximal and minimal displacement
values in atomic chains where the breather has propagated over a
specified computational time interval.  We assign the index $m$ to the
horizontal main chain of atoms along which the breather has
propagated, and indices $m\pm{k}$ where $k=1,2,3$ to the adjacent
chains of atoms, see Fig.\ \ref{fig:OnSiteB}.  We refer to these
chains by $y_{m}$. The displacement plot of maximal and minimal values
is shown in Fig.\ \ref{fig:DisplA}.  The figure confirms that the
largest displacement of atoms is on the main chain $y_{m}$ in the $x$
direction with almost zero displacement in $y$ axis direction.  Figure
\ref{fig:DisplA} is in good agreement with Figs.\ \ref{fig:BDisplA}
and \ref{fig:BDisplB}.  Notice that there is still some displacement
in both axis directions for atoms in adjacent chains $y_{m\pm{3}}$.
This is due to the presence of phonons in the lattice.  Compared to
the breather energy, the phonon energy is very small, as can be seen
in Fig.\ \ref{fig:DisplB}, where we plot the maximal energy of atoms
over time.

Evidently, most of the energy is localized on the atoms on the main
chain $y_{m}$ and rapidly decays along the $y$ axis directions.  Thus
in the $y$ axis direction the breather is localized on around five
atoms while at the same time it is localized on around seven to eight
atoms in the $x$ axis direction, see Fig.\ \ref{fig:BDisplA}.  Closer
inspection of Figs.\ \ref{fig:DisplA} and \ref{fig:DisplB} shows that
maximal and minimal displacements in the $x$ axis directions, as well
as the energy, is symmetric with respect to the adjacent chains
$y_{m\pm{k}}$ where $k=1,2,3$, while maximal and minimal displacements
in the $y$ axis directions are anti-symmetric.  This is because when
the breather propagates on the main chain $y_{m}$, it pushes atoms
{\em away} on adjacent chains.  The maximum displacement away from the
main chain is larger compared to the maximum displacement towards the
main chain $y_{m}$.

The displacement values in Fig.\ \ref{fig:DisplA} are dependent on
the values of $\bar{\epsilon}$ and $\gamma$.  For larger values of
$\gamma$, we observed larger displacement values in the $x$ axis
directions of the atoms on the main chain $y_{m}$.  The same is true
for smaller values of $\bar{\epsilon}$.  Interestingly, smaller values
of $\bar{\epsilon}$ gave smaller values of displacements in the $y$
direction on adjacent chains, and hence increases the
quasi-one-dimensional nature of propagating discrete breathers.

\begin{figure}[ht]
\centering 
\subfigure[]{\label{fig:ChainA}\includegraphics[width=0.32\textwidth]
{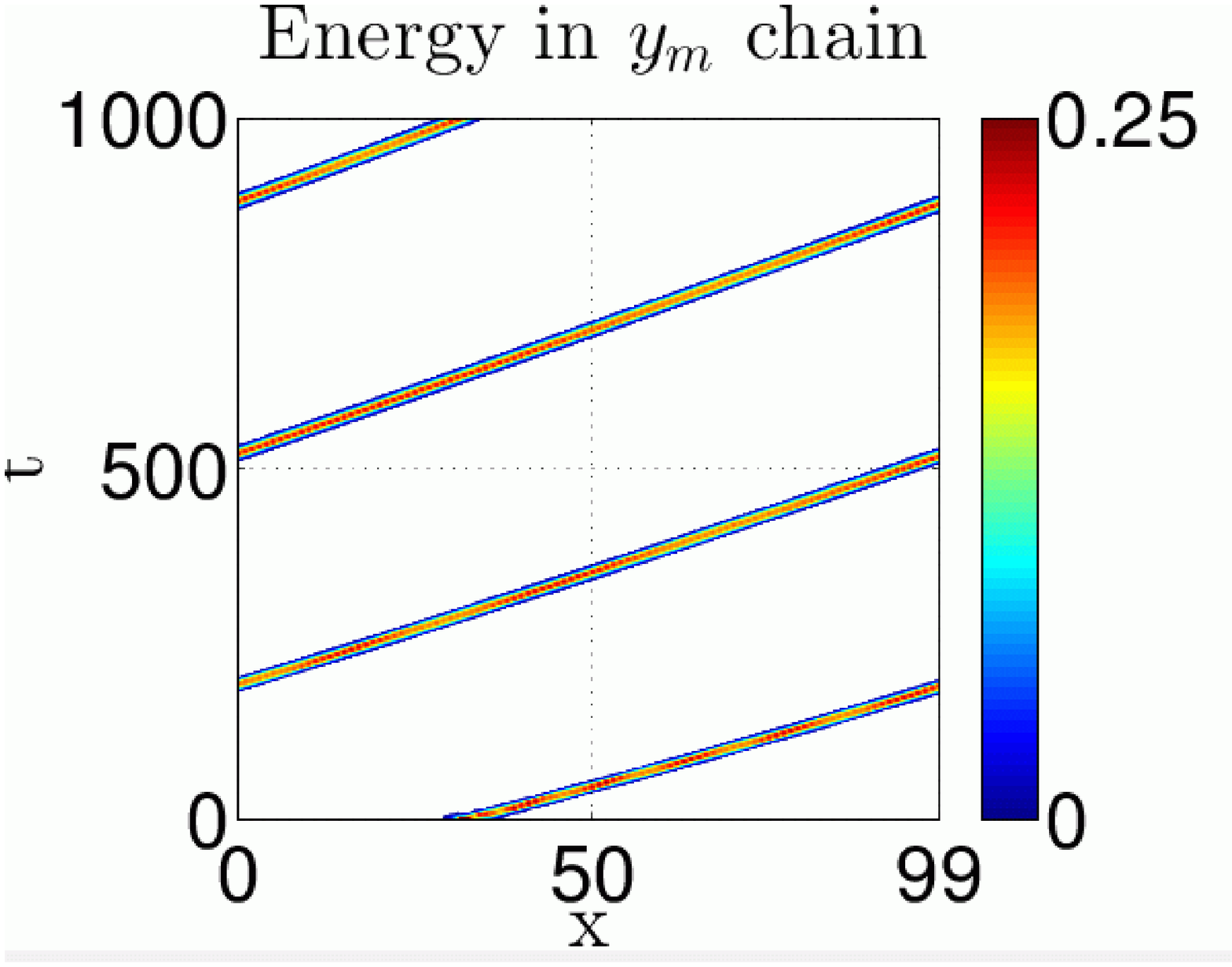}}
\subfigure[]{\label{fig:ChainB}\includegraphics[width=0.32\textwidth]
{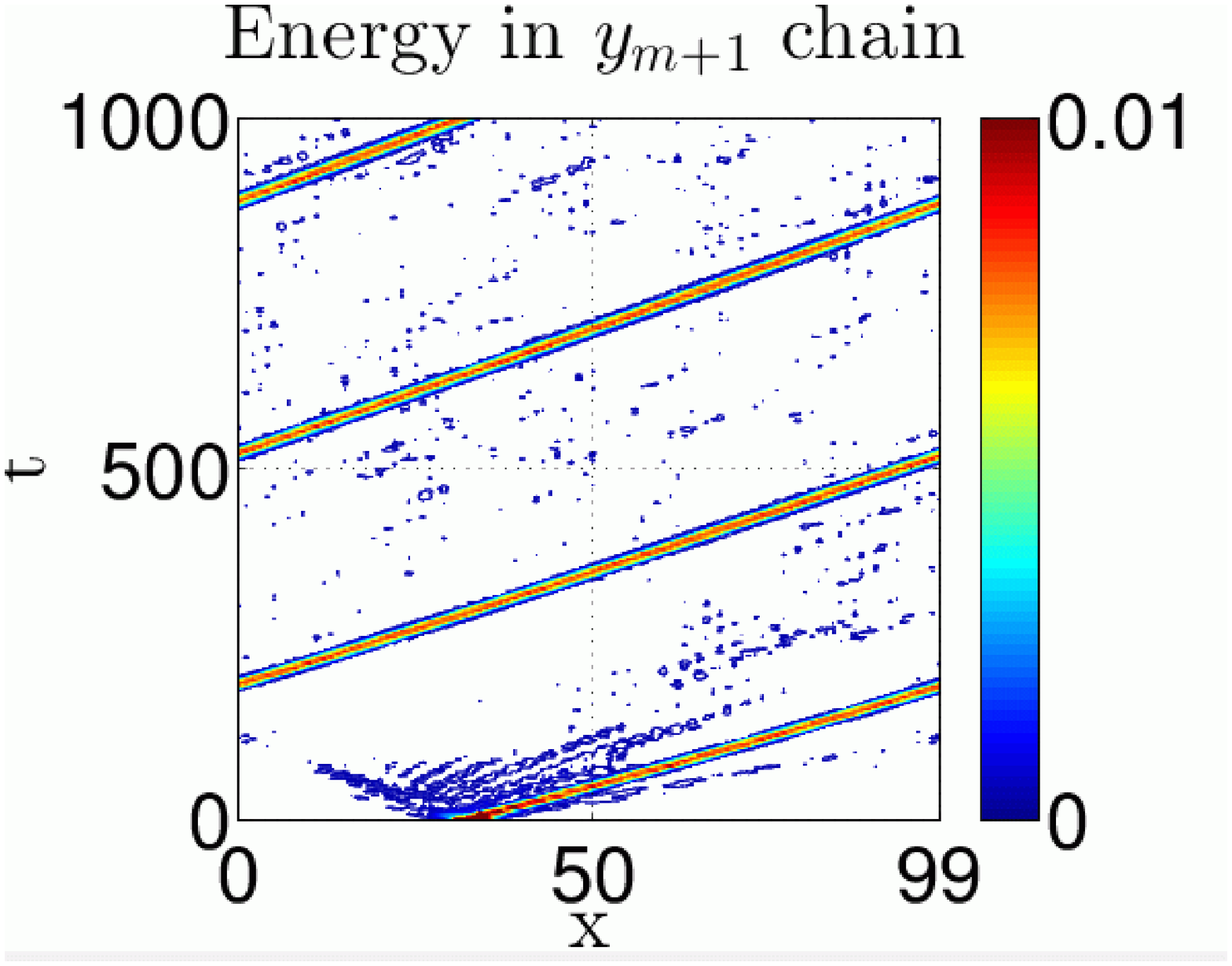}} 
\subfigure[]{\label{fig:ChainC}\includegraphics[width=0.32\textwidth]
{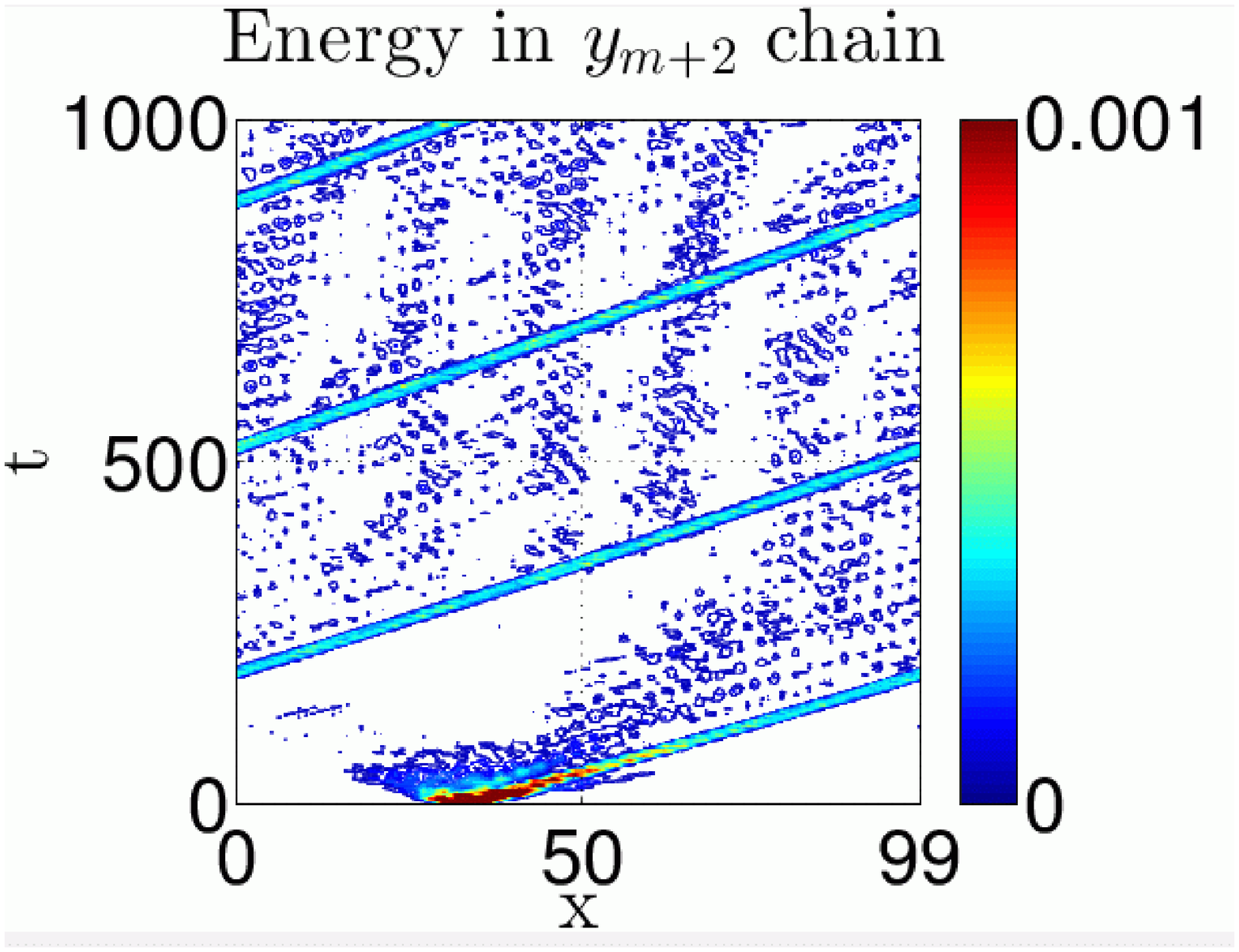}} 
\caption{Contour plots of the breather energy on atomic chains in
  time. (a) breather energy on the main chain $y_{m}$. (b) breather
  energy on the adjacent chain $y_{m+1}$. (c) breather energy on the
  adjacent chain $y_{m+2}$ (note different colour scales).}
\label{fig:ChainEn}
\end{figure}

Indeed, Figure \ref{fig:DisplB} indicates
the quasi-one-dimensional nature of propagating discrete breathers.
Despite the small amount of energy in adjacent chains of atoms, the
energy is still strongly localized.  That can be seen in Fig.\
\ref{fig:ChainEn}.  In Figure \ref{fig:ChainA}, we plot the energy
density function of the chain $y_{m}$ at each time unit.  Similarly we
plot the energy density functions as a function of time in the
adjacent chains $y_{m+1}$ and $y_{m+2}$, see Figs.\ \ref{fig:ChainB}
and \ref{fig:ChainC}, respectively.  As before, for plotting purposes
we interpolated the results on the uniform mesh.  

We pick maximal colour scales according to the values in Fig.\
\ref{fig:DisplB}.  The amount of the breather energy in the chain
$y_{m}$ is much higher as compared to the phonon energies; none of
these energies are visible in Fig.\ \ref{fig:ChainA}. On the contrary,
the small amount of phonon energy is visible in Figs.\
\ref{fig:ChainB} and \ref{fig:ChainC}, thus confirming the presence of
phonon waves in the lattice.  The damping initiated in the simulation
at initial times at the boundaries does not remove the phonons
completely from the system, and these phonons will
affect the long term solution of the propagating discrete breather in
the lattice with periodic boundary conditions.  Questions regarding
the energy loss by the breather solution, its velocity, focusing
properties and lifespan will be addressed in the following sections.

\subsection{Focusing of discrete breathers in frequency space}
\label{sec:Focussing}
\begin{figure}[ht]
\centering 
\subfigure[]{\label{fig:AmplA}\includegraphics[width=0.48\textwidth]
{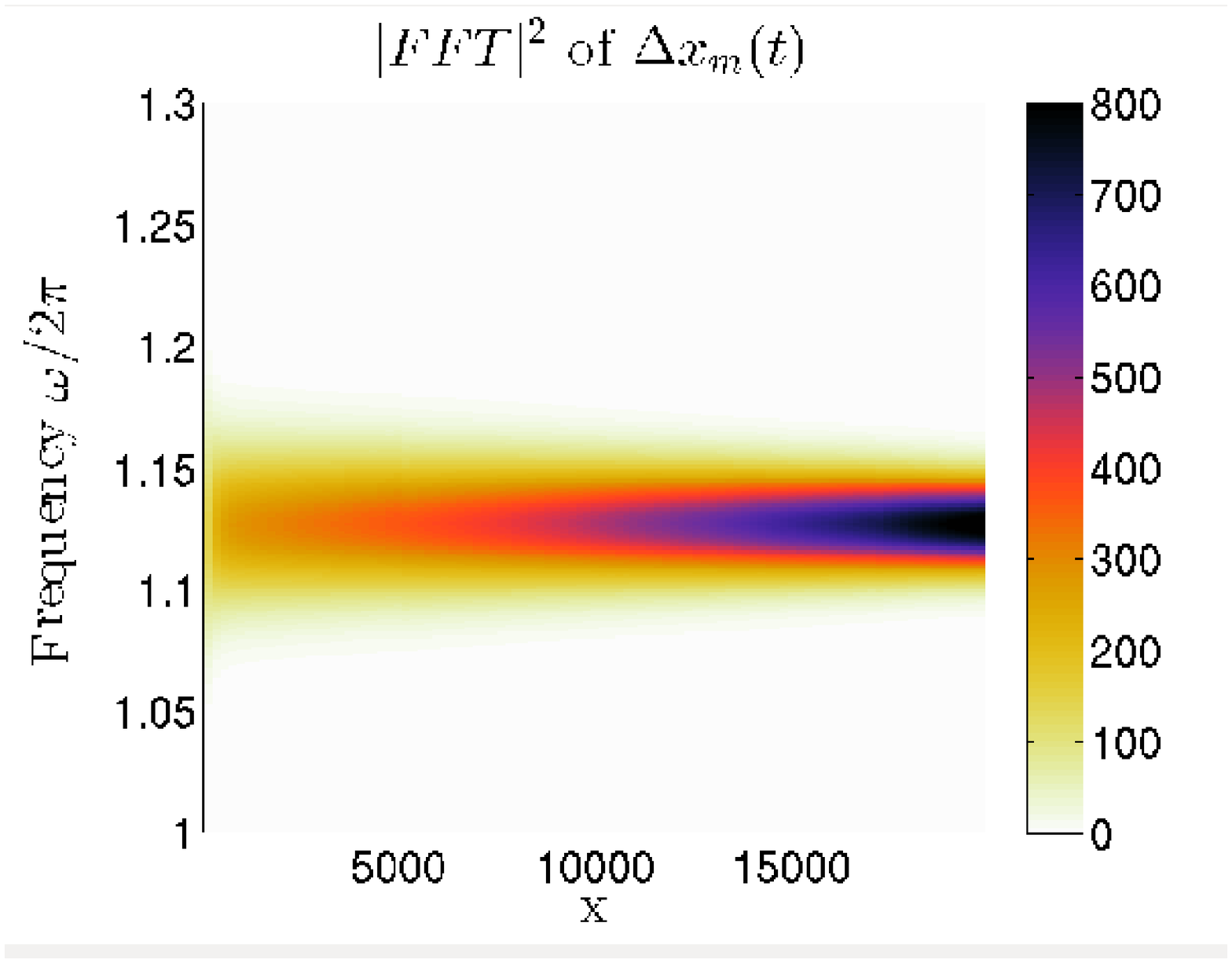}}
\subfigure[]{\label{fig:AmplB}\includegraphics[width=0.48\textwidth]
{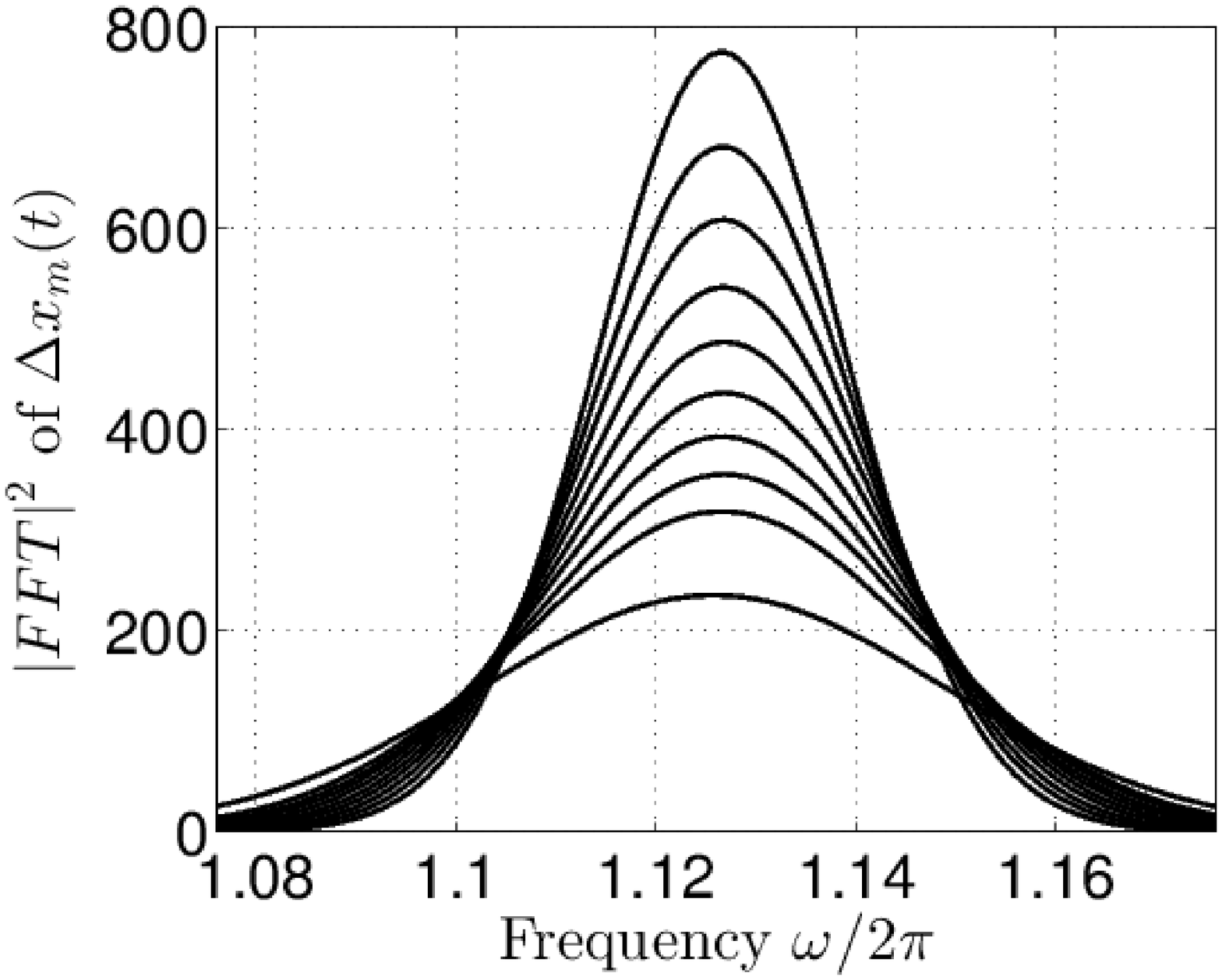}} 
\caption{Propagating breather frequency spectrum. (a) amplitude
  squared of the time series of displacement function
  $\Delta{x}_{m}(t)$ on the main lattice chain $y_{m}$. (b) ten
  cross-sections of plot (a).}\label{fig:Ampl}
\end{figure}

In the previous section, we described how to excite propagating
discrete breathers and discussed the energy localization properties in
atomic chains of the propagating discrete breather.  In this section
we study a novel feature, the focusing property in the frequency
domain, that is, the spectral properties of 2D propagating discrete
breather solutions.  For our study we consider numerical simulations
on the rectangular long strip lattice: $N_{x}=20000$ and $N_{y}=16$,
with periodic boundary conditions.  We integrate in time until the
breather has reached the right hand end of the lattice, that is, after
around $10^5$ time units in our example.  For this example we kept
damping at the upper and lower boundary until the breather has passed
$500$ sites, to remove some amount of the phonons from the lattice,
thus obtaining cleaner data.  We collected time series data of the
displacement function $\Delta{x}_{m}(t)$ at $100$ equally spaced atoms
on the main chain along which the breather propagates.  From the
data obtained, we compute the spectrum and plot the squared amplitude 
of the discrete Fourier transform in Fig.\ \ref{fig:AmplA}.  We
illustrate the same result but in the squared amplitude versus
frequency domain in Fig.\ \ref{fig:AmplB}, where we plot each tenth
section of Fig.\ \ref{fig:AmplA}.

\begin{figure}[ht]
\centering 
\includegraphics[width=0.32\textwidth]{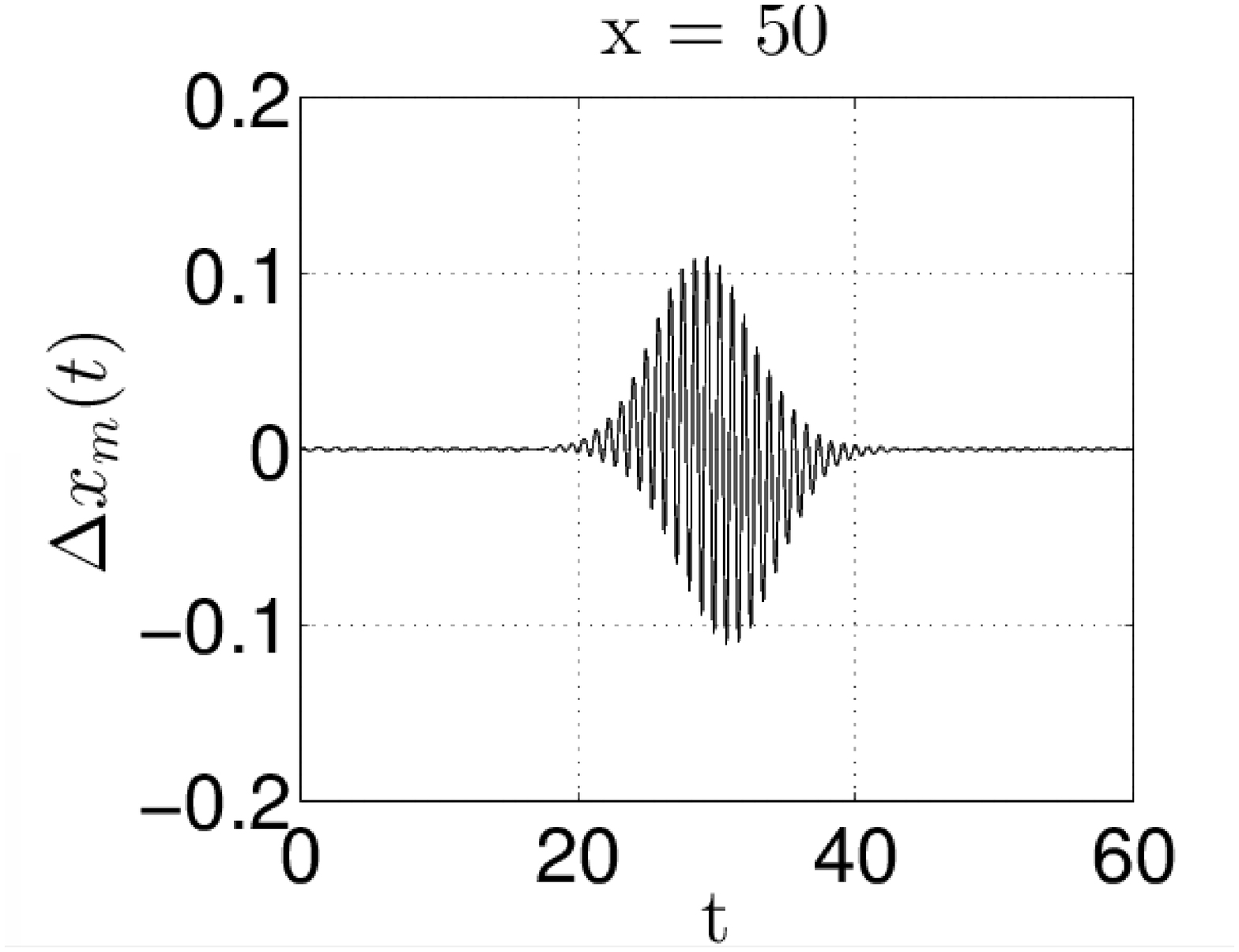}
\includegraphics[width=0.32\textwidth]{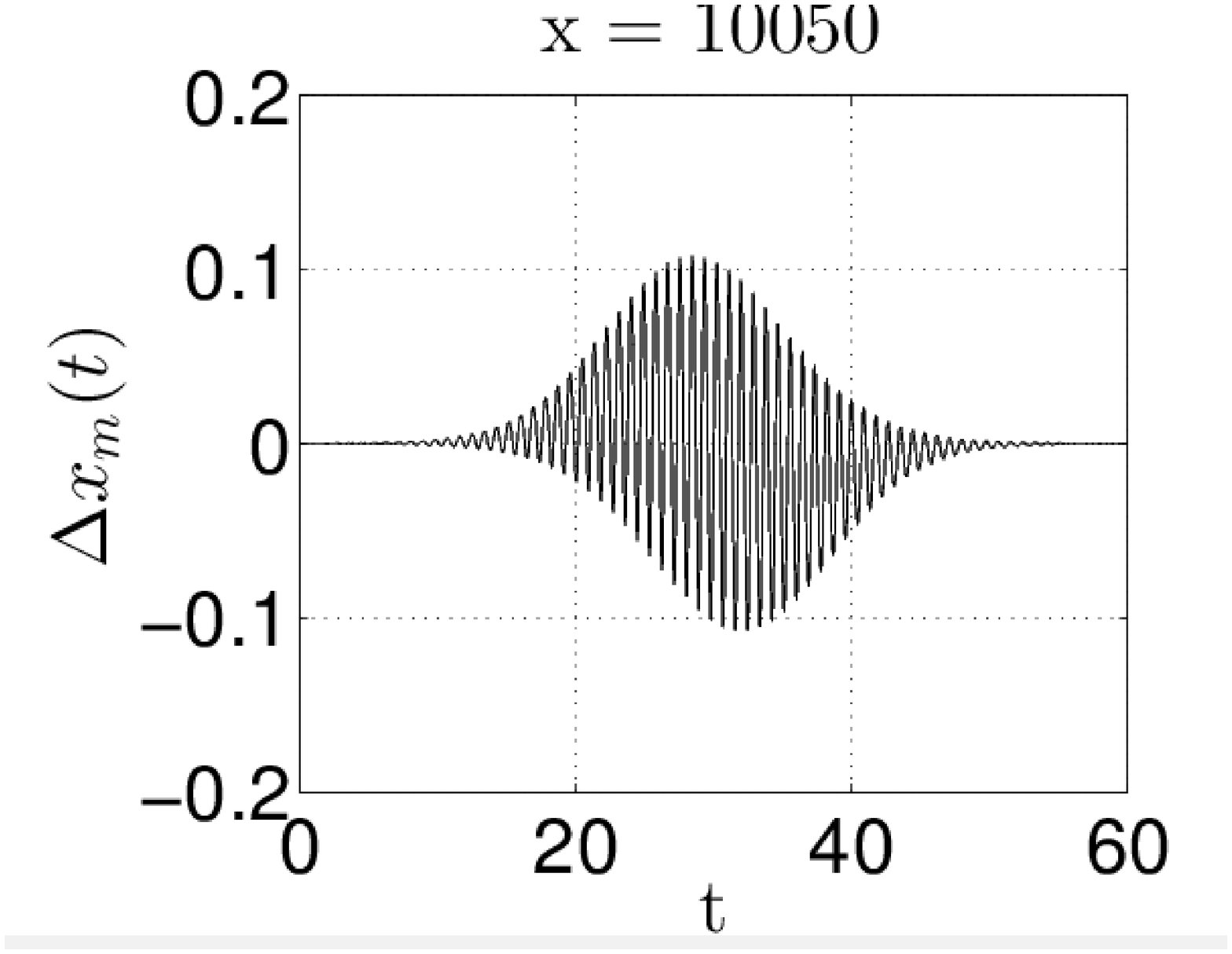} 
\includegraphics[width=0.32\textwidth]{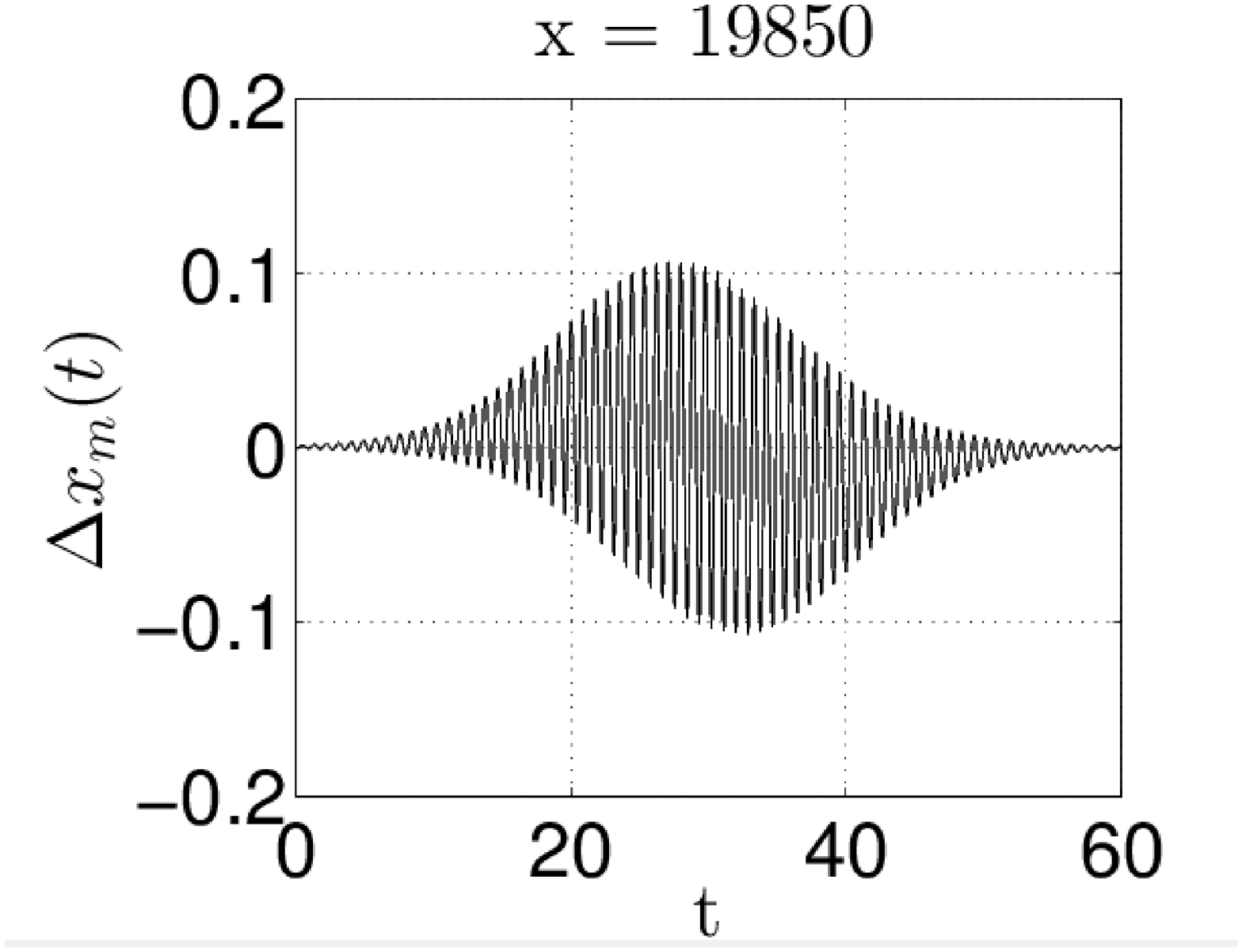} 
\caption{Spreading of the propagating breather solution in time, as
  demonstrated by the displacement function $\Delta{x}_{m}(t)$ at
  three locations in the spacial domain on the chain $y_{m}$.
  Computational times are normalized to the same time
  axis.}\label{fig:BSol}
\end{figure}

Two main observations can be drawn from Fig.\ \ref{fig:Ampl}.  The
first is that the breather frequency is above the phonon frequency
band, compare Fig.\ \ref{fig:AmplA} to Fig.\ \ref{fig:DispRelA}. The
second observation is that the propagating breather is focusing in
frequency space as it evolves. As a result, we observe spreading of
the breather in the time domain, which is illustrated in Fig.\
\ref{fig:BSol}. In Figure \ref{fig:BSol}, we show the time series of
the displacement function $\Delta{x}_{m}(t)$ on a normalized time axis
of three atoms from the main chain $y_{m}$.  Notice that the amplitude
of the breather in Fig.\ \ref{fig:BSol} does not change, only the
width of the wave. Importantly, we observed such focusing effects also
in 1D versions of our 2D lattice model. This naturally raises the
question whether the spreading of the breather occurs also in the spatial
domain. In our numerical results, we did not observe such
phenomenon. It would be interesting to see if such frequency
sharpening, as breathers evolve, also arises in other 1D and 2D
models.

In our long strip lattice simulations we were not able to reach
saturation in frequency space, we need simulations over longer lattice
strips and for longer times.  See the results of
Sec. \ref{sec:LongLived}, where we have performed long time simulation
of propagating breather solution on a $200\times16$ lattice with
periodic boundary conditions.  Notice the relative saturation in the
frequency space, despite the presence of phonons.

\begin{figure}[ht]
\centering 
\subfigure[]{\label{fig:InTimeA}\includegraphics[width=0.48\textwidth]
{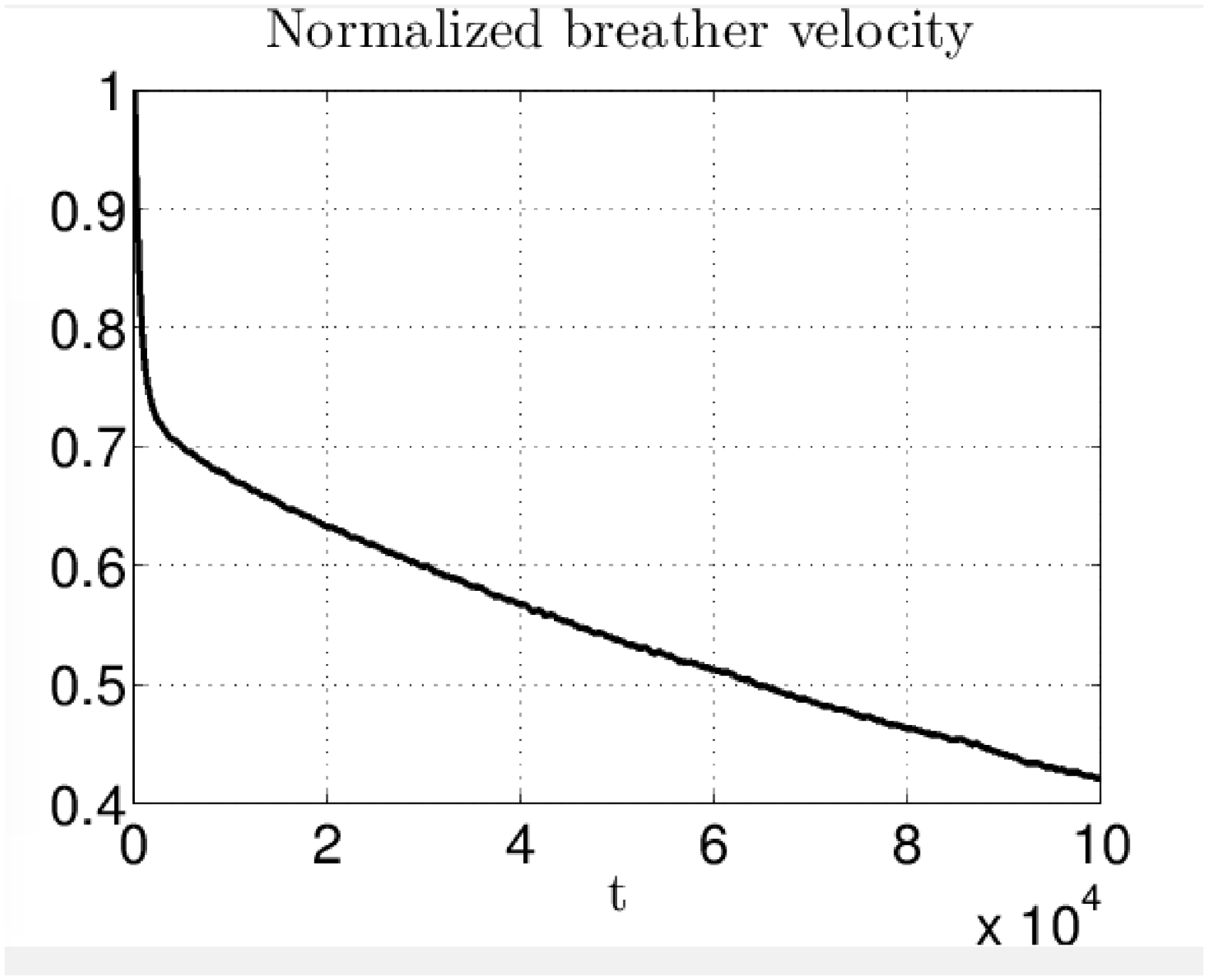}}
\subfigure[]{\label{fig:InTimeB}\includegraphics[width=0.48\textwidth]
{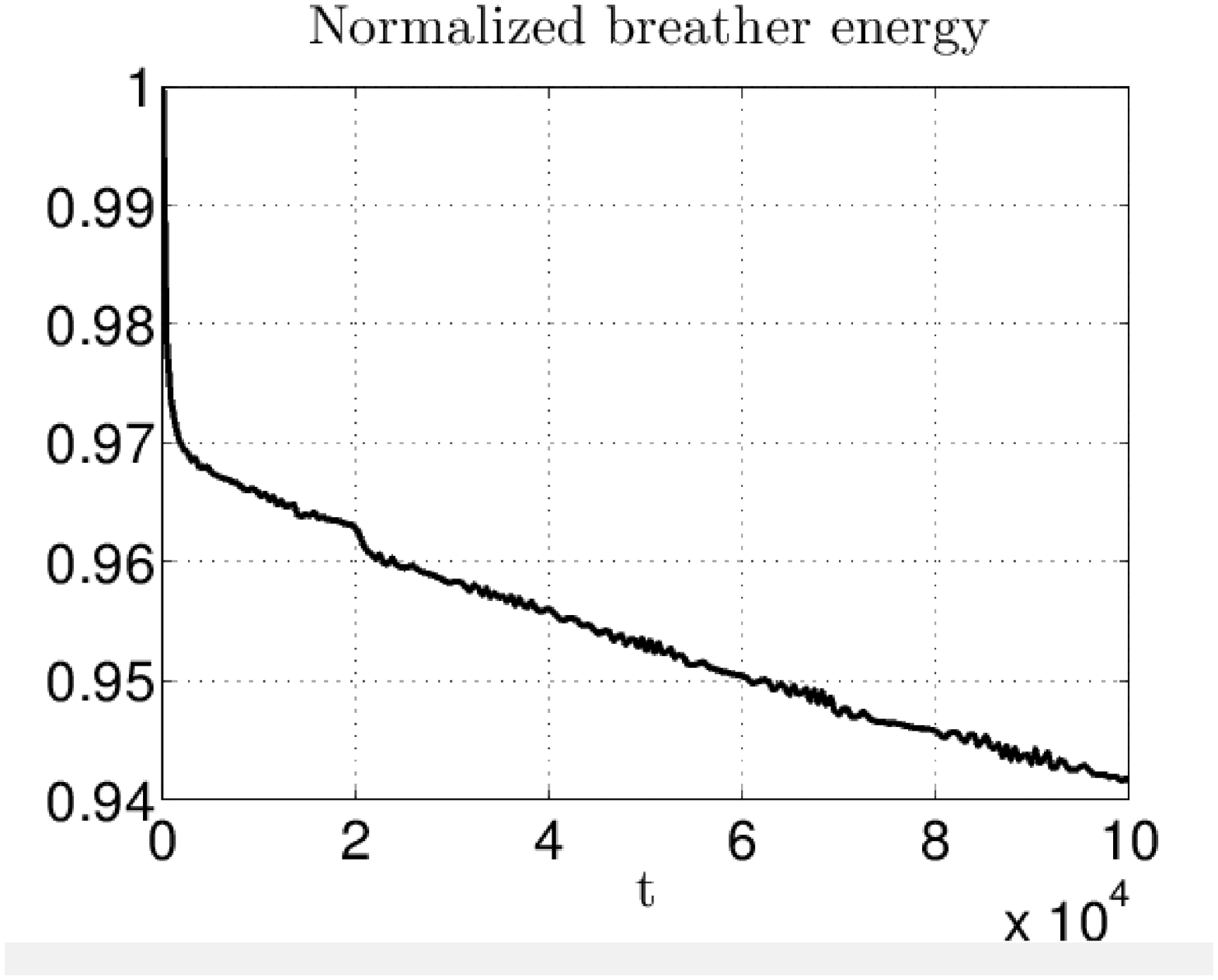}} 
\caption{Long strip lattice simulation, (a) normalized breather
  velocity in time, (b) normalized breather energy in time.}
\label{fig:InTime}
\end{figure}

The spreading of the breather in the time domain suggests that the
breather is slowing down, since it takes longer to pass through one
atom in space.  We confirm this by computing the breather velocity in
time, see Fig.\ \ref{fig:InTimeA}. The graph is obtained by tracking the
location of the breather in space.  From this data we compute the
breather velocity. The smoothed out normalized curve of breather
velocity is shown in Fig.\ \ref{fig:InTimeA}.  At the same time, we
compute the breather energy in time, which is estimated from the sum
of energies over multiple atoms. A smoothed out normalized curve of
the breather energy is illustrated in Fig.\ \ref{fig:InTimeB}.  Figure
\ref{fig:InTime} shows that the propagating breather is slowing down
and loosing its energy.  Our numerical tests showed that there is a
strong correlation between breather velocity and focusing in the
frequency space.  The breather focuses when it slows down and
defocuses when it speeds up.  In some rare cases we could observe time
intervals of constant breather velocity with no focusing or defocusing
in frequency space.  In the next section we show that the same
focusing effect is occurring in adjacent chains.

\subsection{Focusing in adjacent chains}
\label{sec:FocussAdj}
\begin{figure}[ht]
\centering 
\subfigure[]{\label{fig:AmplX1}\includegraphics[width=0.48\textwidth]
{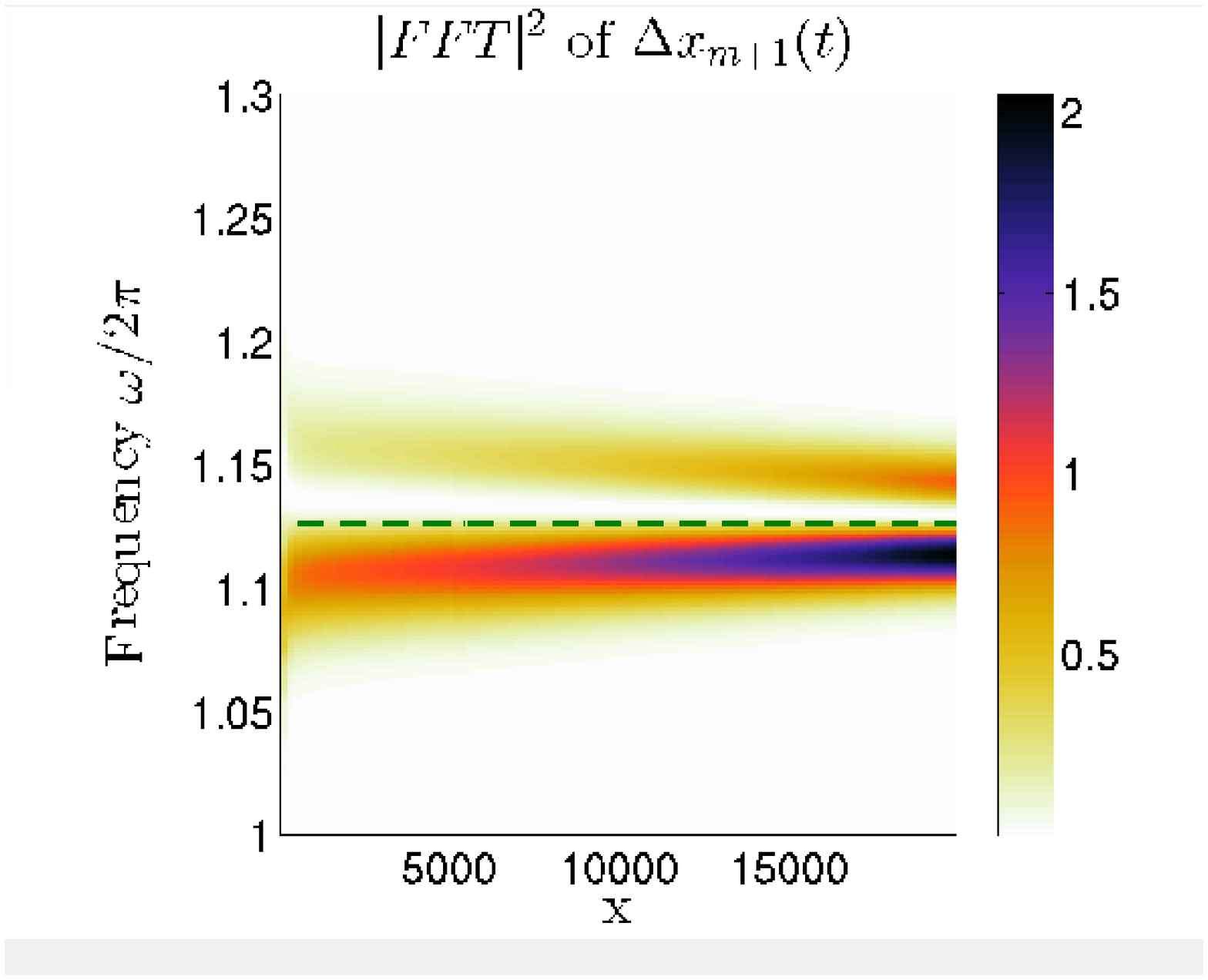}}
\subfigure[]{\label{fig:AmplY1}\includegraphics[width=0.48\textwidth]
{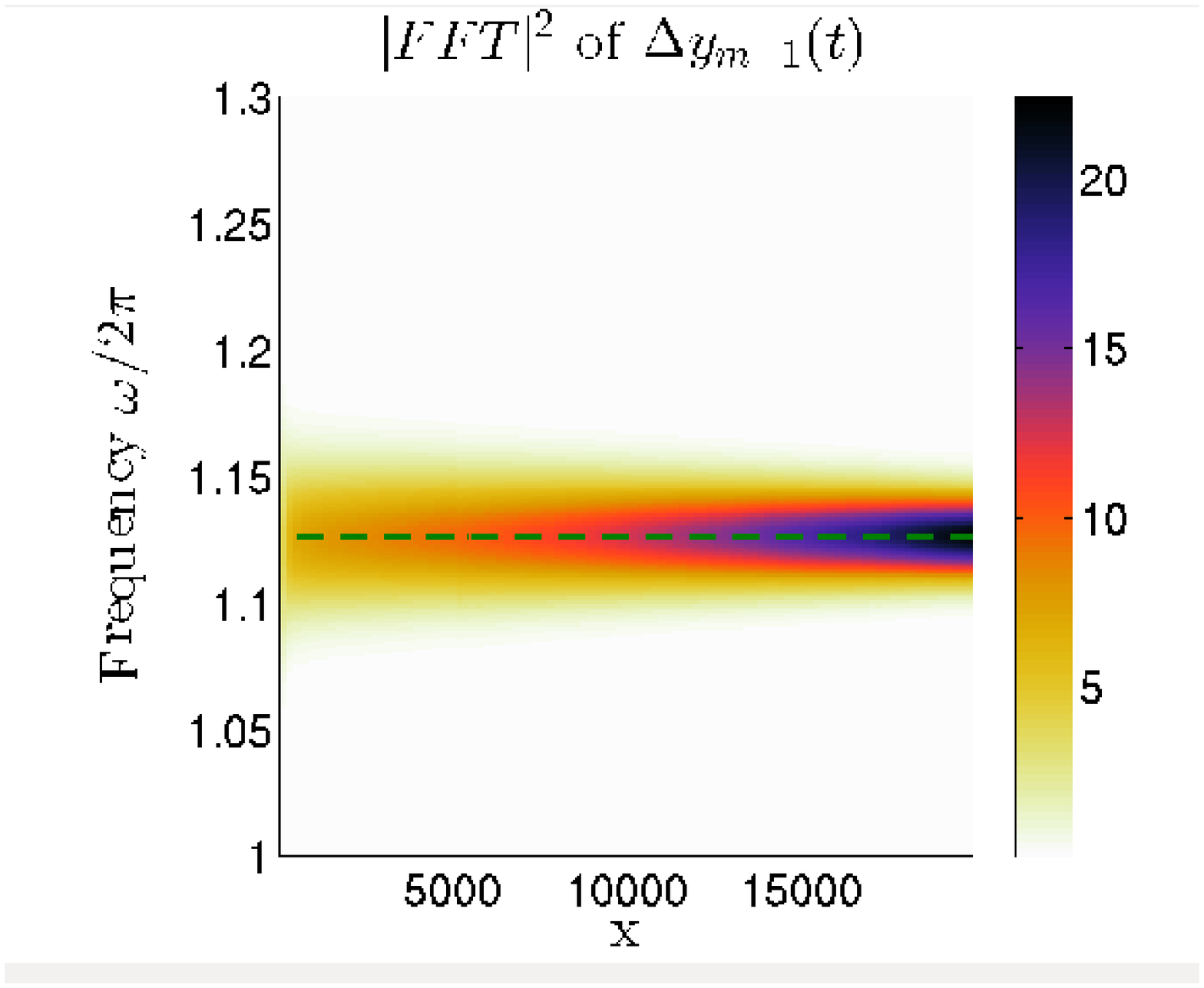}} 
\subfigure[]{\label{fig:AmplX2}\includegraphics[width=0.48\textwidth]
{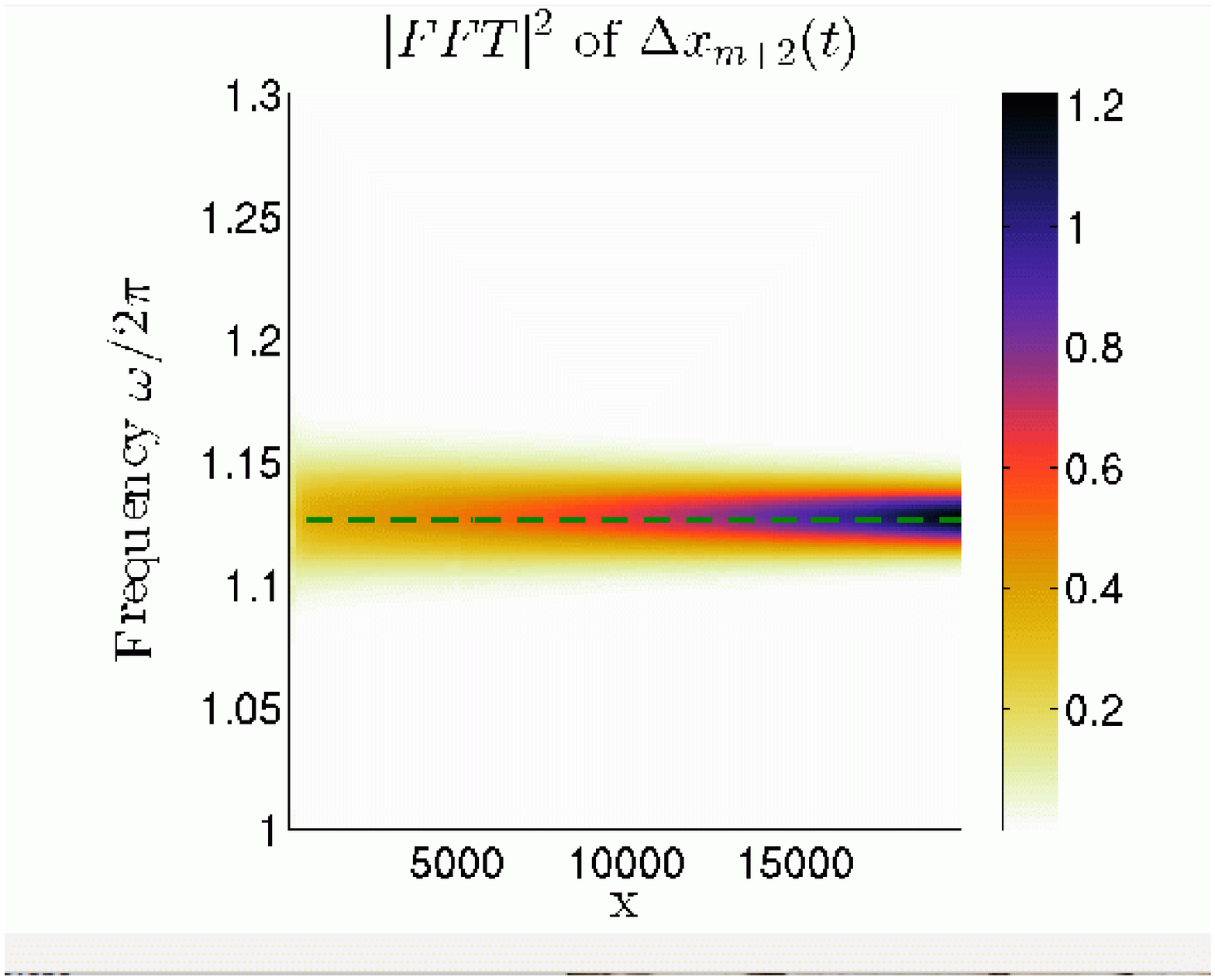}} 
\subfigure[]{\label{fig:AmplY2}\includegraphics[width=0.48\textwidth]
{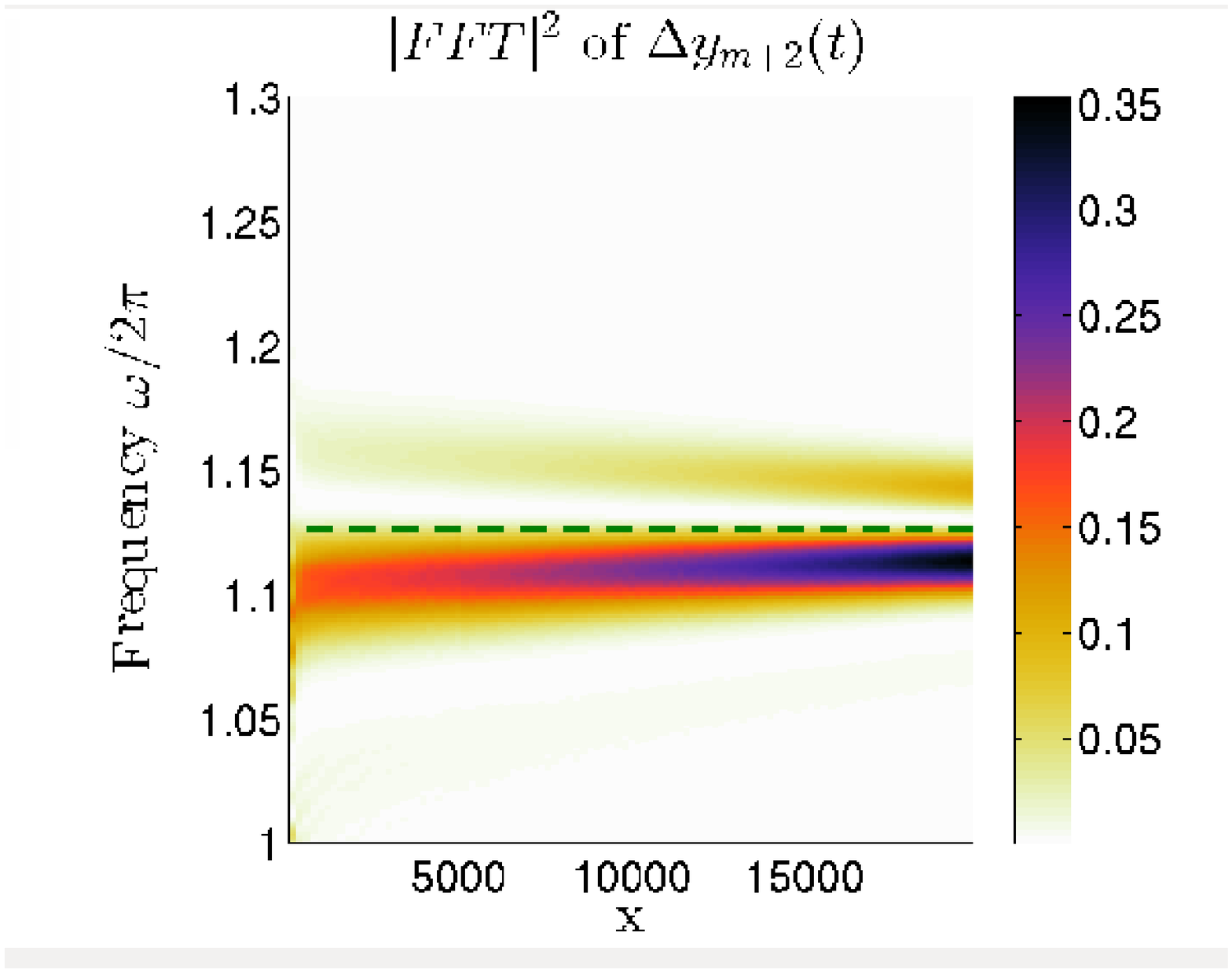}} 
\caption{Frequency spectrum of a propagating breather
    in adjacent chains ($m+1$ and $m+2$). (a) amplitude squared of
  the time series of the displacement function $\Delta{x}_{m+1}(t)$ on
  the lattice chain $y_{m+1}$. (b) amplitude squared of the time
  series of the displacement function $\Delta{y}_{m+1}(t)$ on the
  lattice chain $y_{m+1}$. (c) amplitude squared of the time series of
  the displacement function $\Delta{x}_{m+2}(t)$ on the lattice chain
  $y_{m+2}$. (d) amplitude squared of the time series of the
  displacement function $\Delta{y}_{m+2}(t)$ on the lattice chain
  $y_{m+2}$.}\label{fig:AmplXY}
\end{figure}

In this section we demonstrate the focusing properties of the
propagating breather in adjacent chains of atoms.  We consider the
same numerical example above and collect time series data of
displacement functions $\Delta{x}_{m+1}(t)$, $\Delta{y}_{m+1}(t)$,
$\Delta{x}_{m+2}(t)$ and $\Delta{y}_{m+2}(t)$ of equally spaced atoms
in lattice.  Recall that the breather propagates on the lattice chain
$y_{m}$. For the atom displacement functions on adjacent lines, we
produce equivalent figures to Fig.\ \ref{fig:AmplA}, see Fig.\
\ref{fig:AmplXY}.  The dashed line indicates the dominant frequency of
the Fig.\ \ref{fig:AmplA}.  All plots of Fig.\ \ref{fig:AmplXY} show
the same focusing effect, but with smaller amplitudes as indicated by
the colour bars.

\begin{figure}[ht]
\centering 
\subfigure[]{\label{fig:BSolX1Y1}\includegraphics[width=0.48\textwidth]
{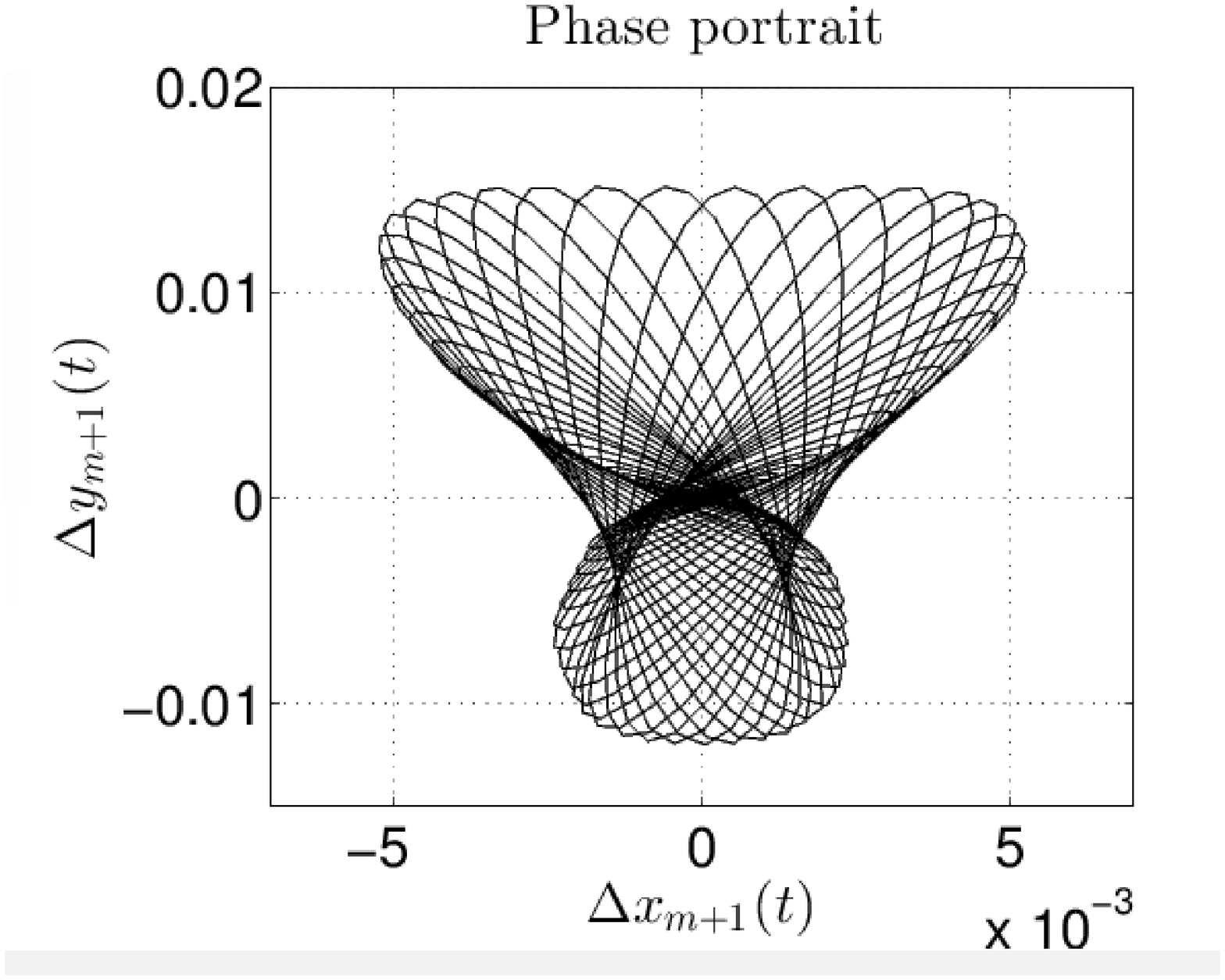}}
\subfigure[]{\label{fig:BSolX2Y2}\includegraphics[width=0.48\textwidth]
{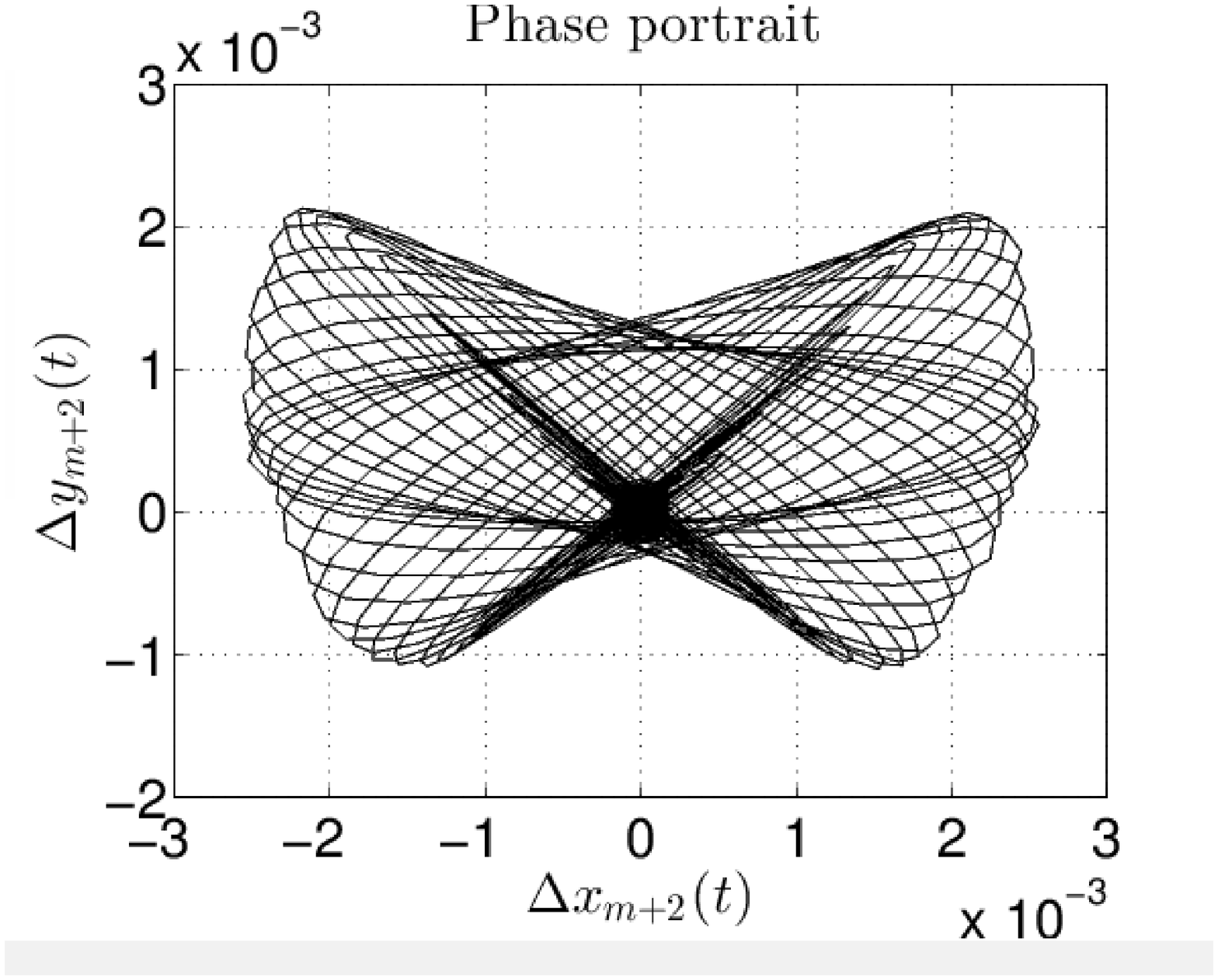}} 
\caption{2D displacements of atoms in lattice chains $y_{m+1}$ and
  $y_{m+2}$ over the time when breather has passed through. (a) phase
  portrait of $\Delta{x}_{m+1}(t)$ and $\Delta{y}_{m+1}(t)$. (b) phase
  portrait of $\Delta{x}_{m+2}(t)$ and
  $\Delta{y}_{m+2}(t)$.}\label{fig:BSolXY3D}
\end{figure}

It is interesting that Figs.\ \ref{fig:AmplY1} and \ref{fig:AmplX2}
show focusing towards the same dominant frequency of the displacement
function $\Delta{x}_{m}(t)$ while the displacement functions
$\Delta{x}_{m+1}(t)$ and $\Delta{y}_{m+2}(t)$ appear to be focusing on
two frequencies above the phonon band, see Figs.\ \ref{fig:AmplX1} and
\ref{fig:AmplY2}, respectively.  This split of frequencies can be
attributed to modulation by the rotational frequencies.  That can be
seen in Fig.\ \ref{fig:BSolXY3D}, where we plot phase portraits of the
displacement functions of one atom in the middle of the computational
domain over the time interval when the breather passes through.  In
other words, Fig.\ \ref{fig:BSolXY3D} shows 2D breather displacements
of atoms in lattice chains adjacent to the main lattice chain $y_{m}$.
In Figure \ref{fig:BSolX1Y1}, we illustrate the phase portrait of the
functions $\Delta{x}_{m+1}(t)$ and $\Delta{y}_{m+1}(t)$, and in Fig.\
\ref{fig:BSolX2Y2} we illustrate the phase portrait of the functions
$\Delta{x}_{m+2}(t)$ and $\Delta{y}_{m+2}(t)$.  Notice the rotational
character of the 2D propagating discrete breather in the adjacent
lattice chains $y_{m+1}$ and $y_{m+2}$.

\clearpage

\subsection{Long-lived breather solutions}
\label{sec:LongLived}
In the sections above, we performed numerical simulations of
propagating discrete breathers and discussed their properties, in
particular, the localization of energy and focusing in frequency
space.  It is still an open question if exact propagating discrete
breathers exist in our model.  Figure \ref{fig:InTime} shows that the
propagating breather slows down and loses its energy.  Long time
studies of propagating discrete breathers are subject to the chaotic
nature of molecular dynamics model, round-off errors and interactions
with the phonon background.  All these aspects lead to the
unpredictable nature of results and sensitive dependence on initial
conditions.  Thus analytical studies are needed to answer the question
regarding existence of propagating discrete breathers.  At the same
time this serves as a good motivation to study breather interactions
with phonons, i.e.~breather propagation in thermalized crystals, and
interactions between breathers themselves, which we report elsewhere.

To contribute to the discussion of the existence of the breather
solutions, we perform a conceptual numerical study of breather
lifespan in a periodic lattice.  We consider a lattice: $N_{x}=200$
and $N_{y}=16$, with periodic boundary conditions.  We excite a
breather solution with atomic momenta pattern (\ref{Pattern}),
$\gamma=0.5$, in the $(1,0)^{T}$ crystallographic lattice direction on
a chain $y_{m}$.  This initial condition is integrated in time until
the breather has passed one million lattice sites, i.e.~crossed the
computational domain $5000$ times.  In our example, that took less
than $10^{7}$ time units.  As before, we kept damping at the upper and
lower boundaries, in this case until the breather has passed $2000$
sites. 

\begin{figure}[ht]
\centering 
\subfigure[]{\label{fig:LongA}\includegraphics[width=0.45\textwidth]
{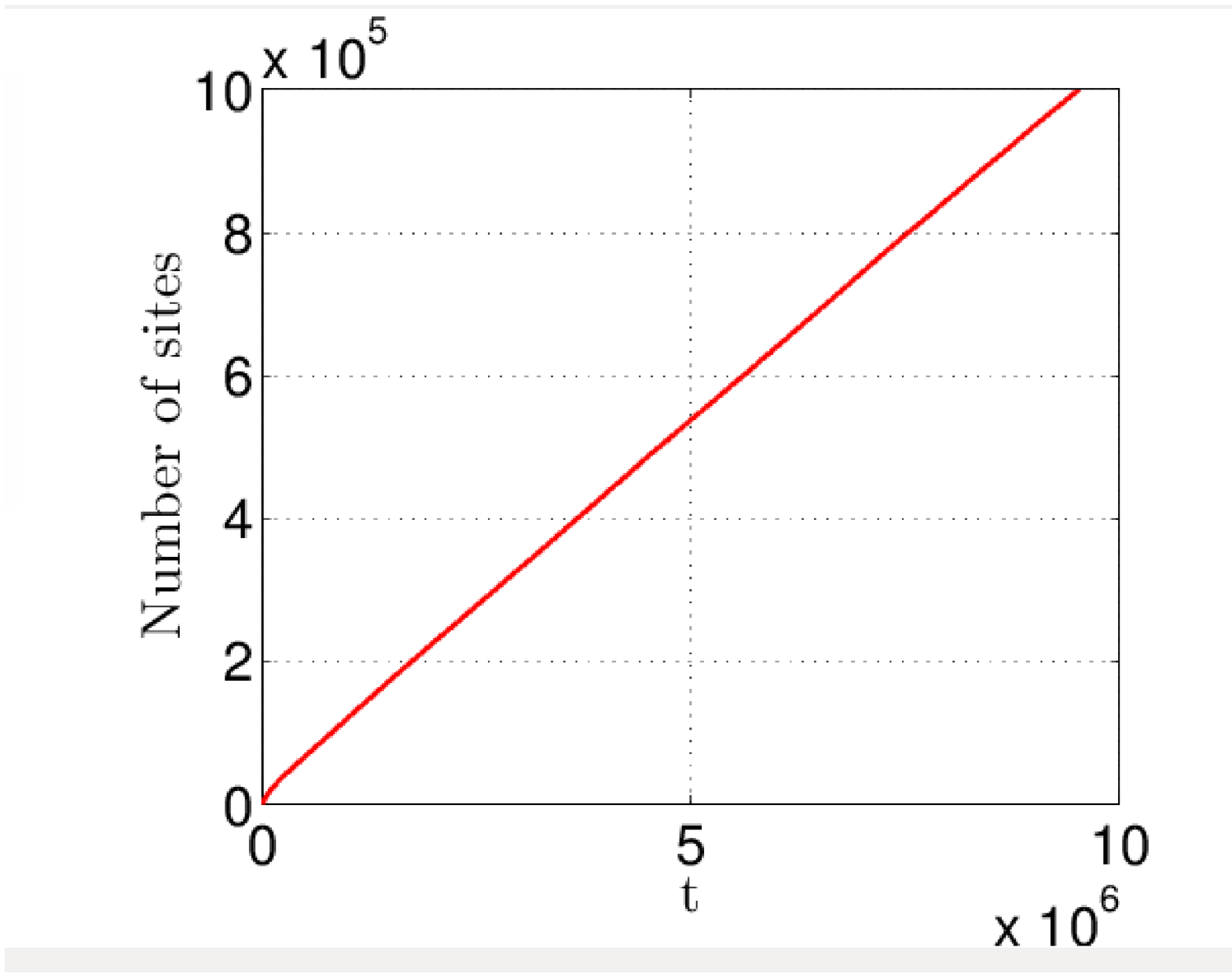}}
\subfigure[]{\label{fig:LongB}\includegraphics[width=0.45\textwidth]
{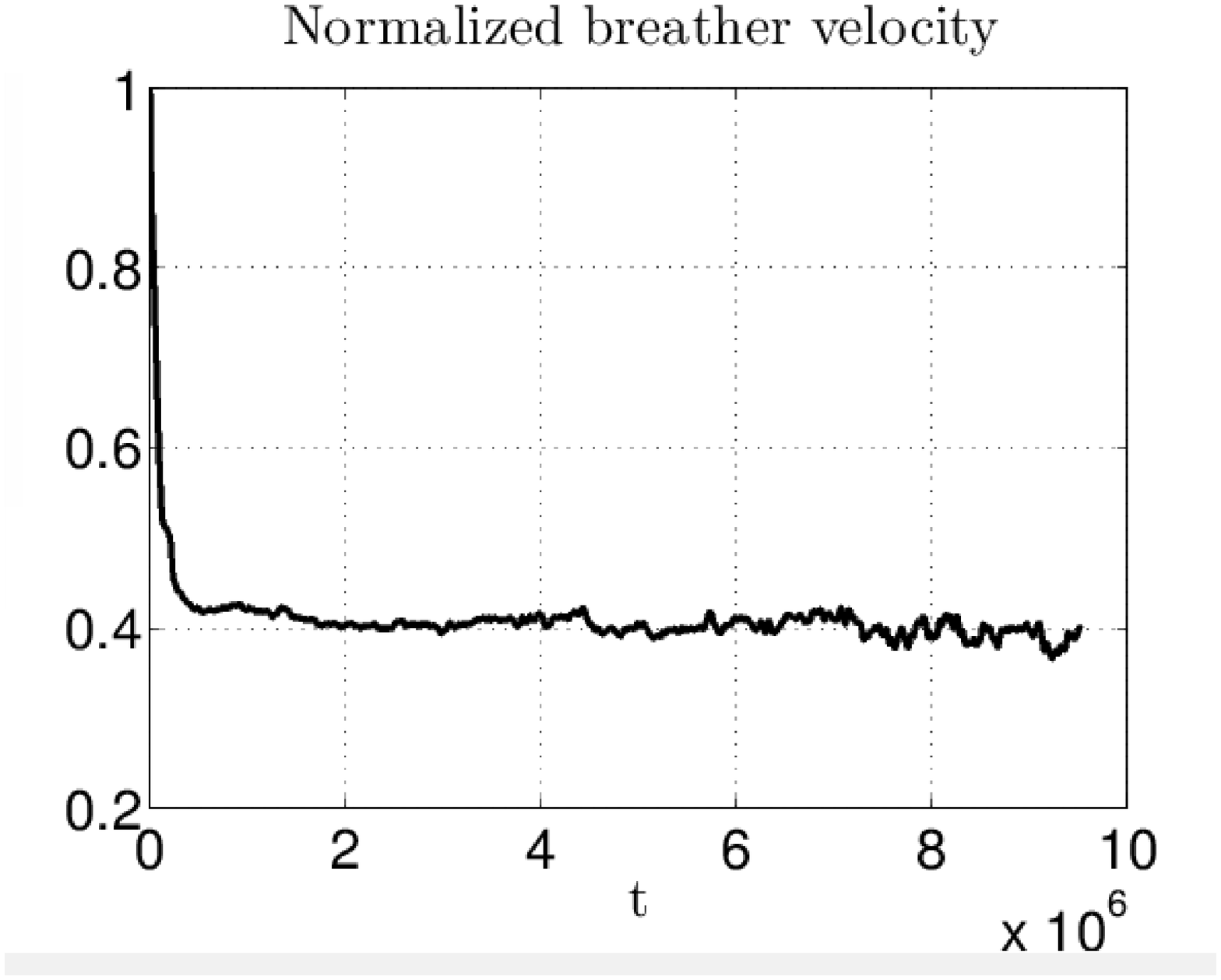}}
\subfigure[]{\label{fig:LongC}\includegraphics[width=0.45\textwidth]
{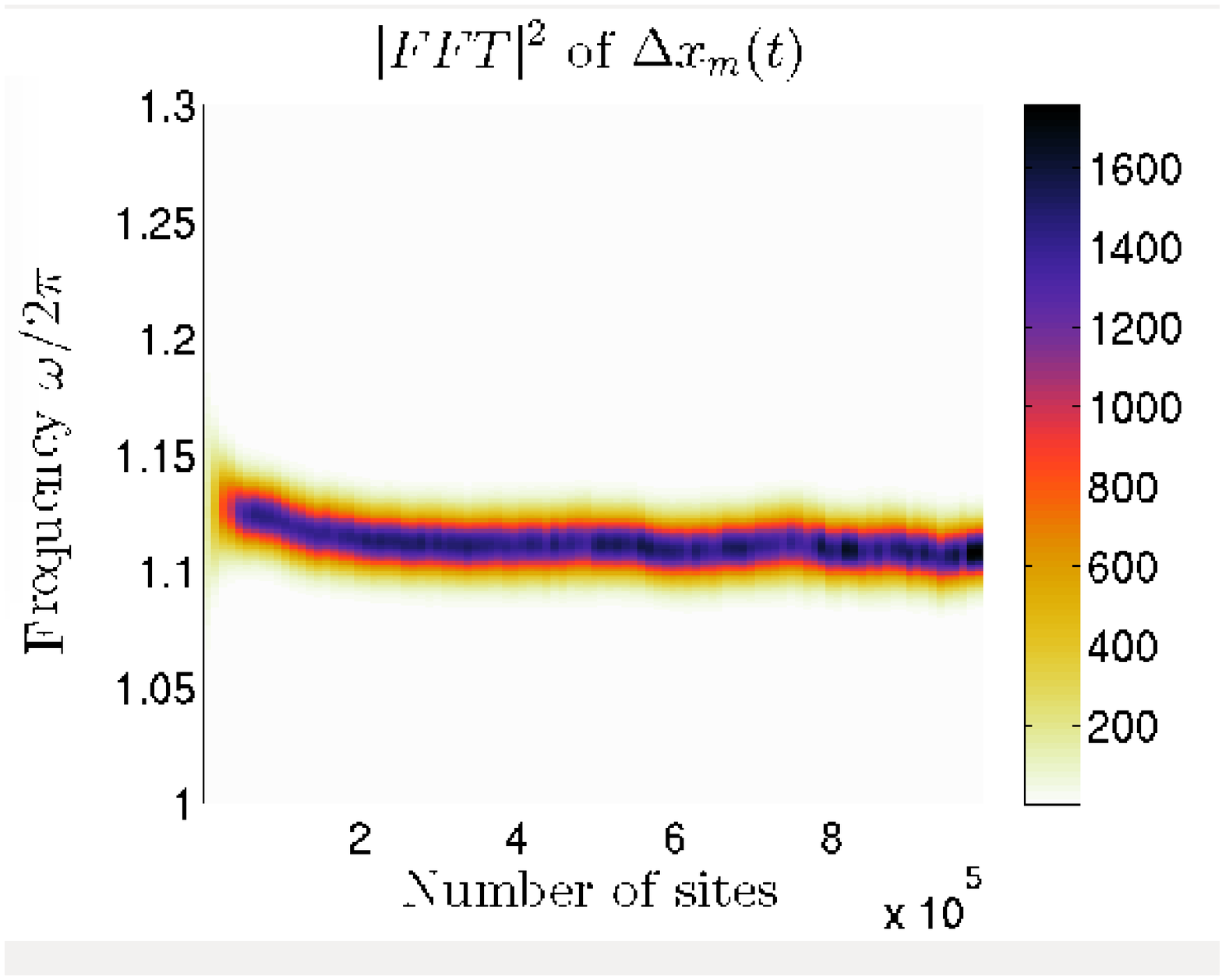}}
\caption{Long-lived breather simulation, $N_{x}=200$ and
  $N_{y}=16$. (a) number of sites the breather has passed versus
  time. (b) normalized breather velocity in time. (c) breather
  frequency spectrum of the displacement function $\Delta{x}_{m}(t)$
  on the $y_{m}$ chain.}
\label{fig:LongL}
\end{figure}

In Figure \ref{fig:LongA}, we plot the number of sites the breather
has passed versus time.  The normalized breather velocity is plotted
in time in Figure \ref{fig:LongB}, computed from the curve in Fig.\
\ref{fig:LongA}.  In addition, after each $50$ breather propagation
cycles, we collected the time series of displacement function
$\Delta{x}_{m}(t)$ of an atom at the middle of the computational
domain on the lattice chain $y_{m}$.  The computed frequency spectrum
of time series is illustrated in Fig.\ \ref{fig:LongC}, where the $x$
axis of the figure refers to the number of sites the breather has
passed in Fig.\ \ref{fig:LongA}.  Figures \ref{fig:LongB} and
\ref{fig:LongC} show the relative saturation in breather velocity and
frequency spectrum over time.  Small scale variations are attributed
to the presence of the weak phonon background.

\clearpage

The numerical results in Figure \ref{fig:LongL} demonstrates
long-lived mobile breather solution in a weakly thermalized background
by phonons. The breather solution has propagated over one million
lattice sites, i.e.\ more than a factor $100\times$ of that reported
by Mar\'in et al.\ \cite{MaEiRu98}.  It would be interesting to
establish a relation between the breather's lifespan and the lattice
temperature for different values of $\bar{\epsilon}$.

\section{Kink solutions}\label{sec:Kinks}
In this section we briefly discuss kink solutions.  Recall that our
model allows atoms to be displaced outside unit cells, in comparison
to the model by Mar\'in et al.\ \cite{MaEiRu98}.  To excite kink
solutions, we consider one atom momentum initial kicks
$\B{v}_{0}=\gamma$ in $(1,0)^{T}$ in a crystallographic lattice
direction on the lattice chain $y_{m}$.  We were not able to excite
kink solutions with parameter $\bar{\epsilon}=0.05$ and $\tau=0.04$
values.  By reducing the parameter values of $\bar{\epsilon}$ and
$\tau$ we were able to observe a highly localized short-lived kink
propagating initially along the atomic chain $y_{m}$.

\begin{figure}[ht]
\centering 
\includegraphics[width=0.95\textwidth]{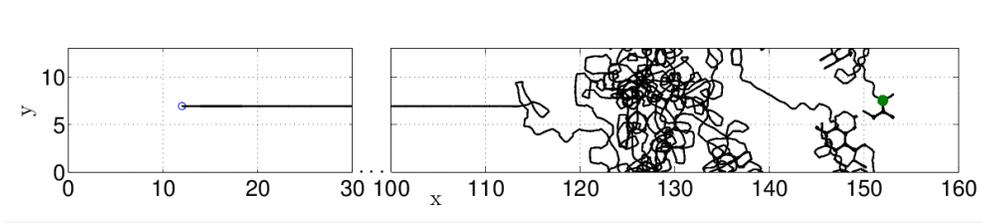}
\caption{Kink solution, position trace  of the maximal energy density
  function. $N_{x}=160$, $N_{y}=16$, $T_{end}=10^4$,
  $r_{cut}=\sqrt{3}$, $\bar{\epsilon}=0.01$, $\B{v}_{0}=4$ and
  $\tau=0.01$.}
\label{fig:KinkSol}
\end{figure}

In Figure \ref{fig:KinkSol}, we show the position of the kink, as
estimated by the position of the maximal energy density, as it
evolves.  It starts at the hollow circle on the left, and initially
travels in a straight line along a crystallographic lattice direction.
We suppress plots of other lower energy excitations created by the
kick, such as breathers, phonons, etc.  The kink radiates energy
through phonons and eventually switches to a more random route,
eventually coming to a halt at the position marked by a filled circle
on the right.  For topological reasons the kink cannot be destroyed
unless it collides with an anti-kink.

We have performed numerical tests with different values of initial
kicks $\B{v}_{0}$, $\bar{\epsilon}$, $r_{cut}$ and time step $\tau$,
and did not observe solutions which persisted along a single
direction, in contrast to \cite{bel14a}, where the same egg-box carton
on-site potential (\ref{OnSiteFunc}) was considered, but used a
piecewise polynomial interaction potential instead of the
Lennard-Jones (\ref{LJpot}).  In this latter paper with this different
potential, we did observe long-lived kink solutions.  These findings
suggest that the choice of interaction potential and its relative
strength with respect to the on-site potential may play a significant
role in the stability of kink solutions. It remains to be seen if
existing or novel 2D materials can exhibit such kinks in physical
situations.
   
\section{Summary}\label{sec:Summary}
In this article, we provide a detailed qualitative study of
propagation breather solutions, building on and expanding the work of
of Mar\'in et al.\ \cite{MaEiRu98} and performed a qualitative study
of propagating breather solutions.  By considering a periodic smooth
``egg-box carton'' on-site potential, and a scaled Lennard-Jones
interaction potential with a cut-off radius, we have derived a
dimensionless system of equations with one dimensionless parameter
$\bar{\epsilon}$.  This parameter is the ratio of the depth of the
interaction potential to that of the on-site potential.  We have found
a range of parameter $\bar{\epsilon}$ values with which mobile
breathers can be observed. This parameter range was found to be in a
good agreement with numerical observations and the considerations of
\cite{MaEiRu98} that both potentials should be of equal relative
strength.

The Lennard-Jones interaction potential with cut-off was constructed
such that harmonic approximation agrees with the harmonic
approximation of the Lennard-Jones potential itself.  With this in
mind, we derived the nearest neighbour linearised equations of
phonons. The derivation of the linearised equations and its dispersion
relation allow us to confirm that breather internal frequency spectrum
is above the phonon band.  Thus, together with numerical observations,
we were able to confirm that the propagating localized modes are
optical breather solutions.  In addition, we argue that the linear
nearest neighbour interaction model, together with egg-box on-site
potential could be a suitable model of lower level of complexity for
analytical investigations.  Such an argument follows from a natural
analogy with the 1D discrete sine-Gordon equation.

In the study of propagating breather solutions, we confirmed its
quasi-one-dimensional nature as well as its 2D characteristics.  We
showed that the most of the breather energy is localized on the main
chain of atoms along which the breather propagates, and that it
propagates in crystallographic lattice directions.  In addition, we
showed that there is also a strong localization of energy present in
adjacent chains of atoms.  From the time series of atomic
displacements in both the $x$ and $y$ axis directions, we were able to
demonstrate the 2D rotational character of the atomic motion in
adjacent chains over the time interval when breather passes through.
From the same time series data, we computed the frequency spectrum and
presented the novel finding of breather localisation in frequency
space as it evolves.  This behaviour causes the breather to spread in
time, while preserving its amplitude.  We found a correlation between
the localization in frequency space and the breather's velocity.  We
found the same localization property in adjacent chains of atoms. It
would be very interesting to see if other 1D or 2D models with
breather solutions support the same localization property in frequency
space as the breather evolves.

To reach the saturation regime where the frequency sharpening
stabilises would take a very long time with long strip lattice
simulations, a computationally challenging task.  In addition, the
chaotic nature of the molecular dynamics system, together with the
existent phonon background added by the initial condition, give
further challenges to long time simulations.  We chose to add a small
amount of damping for short initial time interval to reduce the phonon
density in the system.  

To contribute to the open question of whether propagating breather
solutions exist, we performed conceptual long time simulation on a
{\em small} lattice with periodic boundary conditions.  Despite the
presence of the phonon background, we were able to observe a
long-lived breather solution travelling over one million lattice
sites, a factor of $100$ more than previous results.  This numerical
experiment showed a relative saturation in the breather's velocity,
that is, an almost constant velocity with small variations, and a
relative saturation in the frequency spectrum.  The fluctuations in
the velocity and spectral results are due to the presence of the
weak phonon background.  Importantly, this numerical result serves as
a good motivation for a study of breather interactions and propagation
in thermalized lattices, hinting to the possibility of long-lived
breather solutions in more realistic physical situations.

We concluded our findings with a brief discussion of kink solutions.
Our model allows the displacements of atoms out of the unit cell, in
comparison to the model by Mar\'in et al.\ \cite{MaEiRu98}.  We found
no kink solutions travelling long distances, in contrast to the
findings in \cite{bel14a}, where long-lived kink solutions were
observed in a 2D hexagonal crystal lattice with a different
inter-particle potential.  It is clear that kink solutions are
strongly affected by the choice of interaction potentials, and it
remains to be shown if materials exist which can exhibit such kinks in
physical situations.

\section*{Acknowledgements}
JB and BJL acknowledge the support of the Engineering and Physical
Sciences Research Council which has funded this work as part of the
Numerical Algorithms and Intelligent Software Centre under Grant
EP/G036136/1.

\appendix
\section{Cut-off coefficients}\label{App:A}
In this Appendix we derive a linear system of equations and its
solution to find the cut-off coefficients $A_{j}$ for the interaction
potential (\ref{CutPot}).  The cut-off coefficients are determined from
the conditions (\ref{Cond}) which form a linear system of equations:
\begin{equation}\label{App:syst}
\begin{array}{c}
  \displaystyle \sum_{j=0}^{4} \frac{\sigma^{2j}}{r_{c}^{2j}}A_{j} = 0, 
  \quad \sum_{j=1}^{4}\frac{2j\sigma^{2j-1}}{r_{c}^{2j}}A_{j} = 0, 
  \quad \sum_{j=1}^{4}\frac{2j(2j-1)\sigma^{2j-2}}{r_{c}^{2j}}A_{j} = 0, \\
  \displaystyle \sum_{j=0}^{4} A_{j} = -\tilde{V}, 
  \quad \sum_{j=0}^{4} 2jA_{j}  = -r_{c} \tilde{V}_{r},
\end{array}
\end{equation}
where $\tilde{V} = V(r_c)/\epsilon$ and $\tilde{V}_{r}
= \partial_{r}V(r_{c})/\epsilon$.  In addition we require that
$\tilde{V}\to{0}$ and $r_{c}\tilde{V}_{r}\to{0}$ when
$r_{c}\to\infty$.

The system of equations (\ref{App:syst}) can be solved analytically
to give
\begin{align*}
  A_{0} &= \frac{1}{2} {\frac{ \left(r_{c}{\sigma}^{2} -r_{c}^{3}
      \right) \tilde{V}_{r} + \left(8 r_{c}^{2} - 2{\sigma}^{2}
      \right) \tilde{V} }
    { \left( r_{c}^{2}-{\sigma}^{2} \right) ^{4}}\sigma^{6}} , \\
  A_{1} &= -\frac{1}{2} {\frac { \left( 2 r_{c}^{2}{\sigma}^{2} +
        {\sigma}^{4}-3 r_{c}^{4} \right) \tilde{V}_{r} + 24 r_{c}^{3}
      \tilde{V} }
    { \left( r_{c}^{2}-{\sigma}^{2} \right)^{4}}r_{c} {\sigma}^{4}} , \\
  A_{2} &= \frac{3}{2} { \frac{ \left({\sigma}^{4}-\ r_{c}^{4}\right)
      \tilde{V}_{r} + \left( 8 r_{c}^{3} + 4 r_{c} {\sigma}^{2}
      \right) \tilde{V} }
    { \left( r_{c}^{2}-{\sigma}^{2} \right)^{4}}r_{c}^{3}{\sigma}^{2}} , \\
  A_{3} &= -\frac{1}{2} \frac {\left(3{\sigma}^{4} -
      2r_{c}^{2}{\sigma}^{2} -r_{c}^{4}\right)\tilde{V}_{r} +
    \left(8r_{c}^{3}+16r_{c}{\sigma}^{2}\right)
    \tilde{V}}{\left( r_{c}^{2}-{\sigma}^{2} \right)^{4}} r_{c}^{5} , \\
  A_{4} &= \frac{1}{2} \frac { \left(\sigma^{2}-r_{c}^{2}\right)
    \tilde{V}_{r} + 6r_{c}\tilde{V}}{ \left( r_{c}^{2}-{\sigma}^{2}
    \right) ^{4}} r_{c}^{7} .
\end{align*}
The coefficients $A_{j}$ are shown in non-dimensionless form. To
obtain them in dimensionless form, we set $\sigma=1$ and use
dimensionalized functions $\tilde{V}$ and $\tilde{V}_{r}$.

\section{Linear system and outer products of 
direction cosines}
\label{App:B}
In this appendix, we give explicit expressions for the direction cosine
vectors, their outer products and the linear system (\ref{LinSyst}) in
component-wise form.  The atom $(l,m)$ has six neighbouring atoms
$(l+2,m)$, $(l-2,m)$, $(l+1,m+1)$, $(l+1,m-1)$, $(l-1,m+1)$ and
$(l-1,m-1)$, and associated direction cosines $(1,0)^T$, $(-1,0)^T$,
$(1/2,\sqrt{3}/2)^T$, $(1/2,-\sqrt{3}/2)^T$, $(-1/2,\sqrt{3}/2)^T$ and
$(-1/2,-\sqrt{3}/2)^T$.  Thus the six outer products of the direction
cosines are
\[
D_{l\pm{2},m} = \begin{pmatrix} 1 & 0 \\ 0 & 0 \end{pmatrix}, \quad 
D_{l+1,m\pm{1}} = \begin{pmatrix} \frac{1}{4} & \pm\frac{\sqrt{3}}{4}
  \\ 
\pm\frac{\sqrt{3}}{4} & \frac{3}{4}
\end{pmatrix}, \quad D_{l-1,m\pm{1}} = \begin{pmatrix} \frac{1}{4} &
  \mp\frac{\sqrt{3}}{4}
  \\
  \mp\frac{\sqrt{3}}{4} & \frac{3}{4} \end{pmatrix}.
\]
Applying these matrices to (\ref{LinSyst}), we derive linear equations
in component-wise form:
\begin{align*}
\begin{split}
  \ddot{u}_{l,m} =& - 3u_{l,m}+\tfrac{1}{4} \left( u_{l+1,m+1}
    + u_{l+1,m-1} + u_{l-1,m+1} + u_{l-1,m-1} \right)\\
  & + \left(u_{l+2,m} + u_{l-2,m}\right)  \\
  & + \tfrac{\sqrt{3}}{4} \left( v_{l+1,m+1} - v_{l+1,m-1} - v_{l-1,m+}
    + v_{l-1,m-1}\right) - \kappa u_{l,m},
\end{split}
\end{align*}
\begin{align*}
\begin{split}
  \ddot{v}_{l,m} =& - 3v_{l,m}+\frac{3}{4} \left( v_{l+1,m+1}
    + v_{l+1,m-1} + v_{l-1,m+1} + v_{l-1,m-1} \right) \\
  & + \frac{\sqrt{3}}{4} \left( u_{l+1,m+1} - u_{l+1,m-1} -
    u_{l-1,m+1} + u_{l-1,m-1}\right) -\kappa v_{l,m},
\end{split}
\end{align*}
where $\B{w}_{l,m}=(u_{l,m},v_{l,m})$, that is, the displacements of
atom $(l,m)$ from its equilibrium position in the $x$ and $y$
directions.


\end{document}